\documentclass[11pt]{article}

\usepackage[letterpaper,margin=1in]{geometry}
\usepackage{fancyhdr}
\usepackage{titlesec}
\usepackage[most]{tcolorbox}
\usepackage{microtype}

\usepackage[T1]{fontenc}
\usepackage[utf8]{inputenc}
\usepackage{amsmath}

\usepackage{amssymb,amsfonts}
\usepackage{newtxtext,newtxmath}
\newtcolorbox{rqbox}{colback=gray!10!white,colframe=black!80!white,boxrule=0.5pt,arc=2pt,left=6pt,right=6pt,top=6pt,bottom=6pt}

\usepackage{graphicx}
\usepackage{subcaption}
\usepackage{booktabs}
\usepackage{multirow}
\usepackage{array}
\usepackage{makecell}
\usepackage{tabularx}
\usepackage{longtable}
\usepackage{circledsteps}
\usepackage{pifont}
\usepackage{etoolbox}
\usepackage{float}
\AtBeginEnvironment{thebibliography}{%
  \raggedright
  \setlength{\emergencystretch}{2em}%
}
\usepackage{algorithm}
\usepackage{algorithmic}

\usepackage[dvipsnames,table,xcdraw]{xcolor}
\usepackage[normalem]{ulem}
\usepackage[numbers,sort&compress]{natbib}
\usepackage{hyperref}

\definecolor{paperblue}{HTML}{663333}
\definecolor{lightgraybox}{HTML}{FAF7F2}
\definecolor{softborder}{HTML}{D9C9C2}
\definecolor{bodygray}{HTML}{555555}

\hypersetup{
    colorlinks=true,
    linkcolor=paperblue,
    citecolor=paperblue,
    urlcolor=paperblue
}

\titleformat{\section}
  {\color{paperblue}\large\bfseries}
  {\thesection.}{0.6em}{}

\titleformat{\subsection}
  {\color{paperblue}\normalsize\bfseries}
  {\thesubsection.}{0.5em}{}

\titleformat{\subsubsection}
  {\color{paperblue}\normalsize\bfseries}
  {\thesubsubsection.}{0.5em}{}

\newcommand{\papertitle}{%
After Talking with 1,000 Personas: Learning Preference-Aligned
Proactive Assistants from Large-Scale Simulated Persona Interactions%
}

\newcommand{\shortpapertitle}{%
Learning Preference-Aligned Proactive Assistants from Large-Scale Simulated Persona Interactions%
}

\newtcolorbox{abstractbox}{
    enhanced,
    breakable,
    colback=lightgraybox,
    colframe=softborder,
    boxrule=0.4pt,
    arc=1.5pt,
    left=12pt,
    right=12pt,
    top=10pt,
    bottom=10pt,
    borderline west={2.2pt}{0pt}{paperblue}
}

\begin{document}

\definecolor{myblue}{RGB}{0,0,0}
\thispagestyle{empty}

\noindent
\rule{\textwidth}{0.8pt}

\vspace{1em}

\noindent
\parbox{\textwidth}{%
    \raggedright
    \color{paperblue}
    \LARGE\bfseries
    \papertitle
}

\vspace{1em}

\noindent
{\bfseries
Ziyi Xuan,
Yiwen Wu,
Zhaoyang Yan,
Vinod Namboodiri,
Yu Yang
\par}

\vspace{0.3em}

\noindent
{\small
Lehigh University, USA
\par}

\vspace{0.3em}

\noindent
{\small
\{zix222, yiw220, zhy223, vin423, yuyang\}@lehigh.edu
\par}

\vspace{0.25em}



\noindent\rule{\textwidth}{0.8pt}

\vspace{0.25em}

\begin{abstractbox}

Smart assistants increasingly act proactively, yet mistimed or intrusive behavior often causes users to lose trust and disable these features. Learning user preferences for proactive assistance is difficult because real-world studies are costly, limited in scale, and rarely capture how preferences change across multiple interaction sessions.
Large language model-based generative agents offer a way to simulate realistic interactions, but existing synthetic datasets remain limited in temporal depth, diverse personas, and multi-dimensional preferences. They also provide little support for transferring population-level insights to individual users under on-device constraints. We present a population-to-individual learning framework for preference-aligned proactive assistants that operates under on-device and privacy constraints.
Our approach uses large-scale interaction simulation with 1,000 diverse personas to learn shared structure in how users express preferences across recurring dimensions such as timing, autonomy, and communication style, providing a strong cold start without relying on real user logs. The assistant then adapts to individual users on device through lightweight activation-based steering driven by simple interaction feedback, without model retraining or cloud-side updates.
We evaluate the framework using controlled simulations with 1,000 simulated personas and a human-subject study with 34 participants. 
Results show improved timing decisions and perceived interaction quality over untuned and direct-response baselines, while on-device activation steering achieves performance comparable to reinforcement learning from human feedback. Participants also report higher satisfaction, trust, and comfort as the assistant adapts over multiple interaction sessions.

\end{abstractbox}
\vspace{0.3em}

\section{Introduction}
Smart assistants are now embedded in everyday life across personal computers, smartphones, smart speakers, wearable devices (such as watches and emerging body-worn form factors), and in-car systems, supporting work, home, health, and mobility tasks through a mix of reactive commands and increasingly proactive behaviors~\cite{berube2024proactive, zhan2024healthcare, yang2025socialmind, yang2025contextagent}. Beyond responding to explicit commands, these systems increasingly act proactively by surfacing reminders, suggesting actions, and initiating interactions on their own. When well aligned with user needs, such behavior can reduce effort and provide timely support. However, poorly aligned proactive actions often feel intrusive rather than helpful, particularly when they occur at inconvenient times, take too much initiative, or use an inappropriate communication style~\cite{bentley2018understanding, cha2020hello, zargham_i_2022}. Repeated negative experiences can quickly erode trust, leading users to ignore suggestions or disable proactive features altogether~\cite{pielot2018dismissed, gavilan2022exploring}.

Whether proactive behavior is perceived as helpful depends strongly on individual user preferences. Users differ in how much initiative they want an assistant to take, when interruptions are acceptable, and how information should be presented, and these expectations vary across activities and contexts. Supporting proactive assistance therefore requires systems that can adapt behavior across repeated interactions, rather than relying on fixed rules or one-size-fits-all defaults. Designing and evaluating such adaptive assistants, however, remains challenging~\cite{wei2021understanding, zargham2022circumstances, oh2024better}. Traditional approaches rely on controlled laboratory studies or small in-the-wild deployments with human participants, which are costly and logistically complex, limiting their scale, diversity, and duration. These constraints are particularly acute on mobile and ubiquitous systems, where continuous sensing and logging raise deployment concerns~\cite{rooksby2019tracking, shklovski2014sensing, ghosh2023privacy}.

Simulation-based platforms partially address these challenges by enabling large-scale and repeatable studies without requiring early real-world deployment. In other domains, such as robotics, simulation is routinely used to train systems and explore failure modes before real-world use. Systems such as OpenSHS~\cite{alshammari2017openshs} and VirtualHome~\cite{puig2018virtualhome} model environments and activities of daily living, allowing researchers to test assistant behaviors in structured settings. More recent approaches use synthetic data and simulated users with preferences to study proactive interactions at scale~\cite{park2023generative, park2024generative, gidea_imwut, gidea_sensys, gao2024large}. However, the usefulness of simulation for preference learning depends on whether synthetic data captures meaningful preference structure and whether simulated users express preferences realistically across repeated interactions.

Existing synthetic datasets remain limited for preference learning in proactive assistants. Many focus on constrained interaction sequences~\cite{cook2012casas}, assume static or weakly structured preferences~\cite{acon96_home_assistant_requests}, or rely on personas that are loosely grounded in realistic daily activities~\cite{miller2017parlai}. As a result, they rarely provide multi-dimensional preference traces across multiple interaction sessions, and they offer limited guidance on how synthetic data can support learning assistants that operate under on-device constraints.
Recent work such as the GIDEA platform~\cite{gidea_imwut, gidea_sensys} demonstrates that large language model (LLM)-based agents can simulate human participants in assistant studies with high semantic fidelity, reproducing realistic routines, preference expressions, and reactions to proactive interventions. This evidence suggests that LLM-driven simulation can provide a proxy population for studying proactive assistant behavior at scale without collecting real user logs (especially during the early stage). However, prior uses of GIDEA primarily focus on reproducing aggregate interaction patterns. They do not explicitly organize preference differences across individuals in a way that supports personalized learning and on-device adaptation.

Building on this insight, we show how LLM-driven simulation can be used not only for evaluation but also as training data for preference-aligned proactive assistants. Our approach is motivated by a simple structure: across diverse users, evaluations of proactive behavior often refer to recurring aspects of the assistant (for example, timing, autonomy, and communication style), while individuals differ in how they weight these aspects and how they express them during daily activities~\cite{SCHIAFFINO2004129, zargham2022circumstances}. This motivates a population-to-individual learning setup: learn shared patterns from a large synthetic population, then adapt efficiently to each user based on small amounts of on-device feedback~\cite{fallah2020metalearning, chen2024proactive}.

To support this, we extend the GIDEA multi-agent platform to generate a large-scale, multi-session synthetic dataset that captures how diverse users express preferences during daily routines. We construct a dataset with 1{,}000 personas drawn from a census-aligned persona pool, each engaged in week-long activity sequences. We annotate interactions using a compact preference schema and validate that the resulting activity distributions align with real-world time-use statistics. This dataset provides structured population-level preference signals without collecting real user logs.
On top of this dataset, we introduce a two-stage population-to-individual learning framework for preference-aligned proactive assistants that operates under on-device and privacy constraints. The framework first trains on the synthetic population to learn shared patterns of preference expression, then adapts to individual users through lightweight activation-based steering that runs entirely on the device. This approach avoids extensive per-user data collection or cloud-based personalization while maintaining alignment with individual preferences.

We evaluate the proposed framework through both simulation-based experiments and a human-subject study. In controlled simulations with 1{,}000 personas, we compare population-level training strategies and individual adaptation methods across seen and unseen activity contexts. The results show that incorporating explicit preference structure improves preference understanding and timing decisions compared with untuned and direct-response baselines, and that activation-based adaptation achieves performance comparable to reinforcement learning from human feedback while running entirely on the device. In an IRB-approved study with \textcolor{myblue}{34} participants using mobile-style scenarios, users more often prefer responses from the personalized assistant and report higher satisfaction, trust, and comfort as the assistant adapts over multi-session interactions. Together, these results demonstrate that training proactive assistants through LLM-driven simulation offers a practical path toward preference-aligned behavior in real-world mobile and ubiquitous settings.

The main contributions are as follows.
\begin{itemize}
    \item We propose using LLM-driven simulation as a practical source of synthetic interactions to train preference-aligned proactive assistants, not just to evaluate them. This offers key advantages for early-stage assistant research: large scale and long time horizons without recruiting participants, repeatable and controllable experiments, and privacy benefits because training can avoid collecting real user logs.
    
    \item \textcolor{myblue}{We introduce a population-to-individual learning pipeline that bridges simulation and real-world personalization: combining category-structured supervised fine-tuning for shared preference learning with lightweight on-device activation steering for individual adaptation.}

    \item \textcolor{myblue}{We construct and validate a synthetic dataset of multi-session proactive interactions from 1{,}000 census-aligned simulated personas, organized by a compact multi-dimensional preference schema.}

    \item \textcolor{myblue}{We evaluate the framework through controlled simulations and an IRB-approved human study, showing improvements in preference understanding, timing decisions, and user-perceived interaction quality.}
\end{itemize}

\section{Related Work}
\subsection{Design Principles for Proactive Smart Assistants} 
A substantial body of HCI research has established design principles for proactive smart assistants, emphasizing that users evaluate proactive behavior holistically across multiple dimensions. 
Research on mobile notifications shows that receptivity cannot be reduced to interruptibility alone, but depends on the user's ongoing activity, the content and source of an intervention, and prior patterns of engagement~\cite{mehrotra2015designing, pielot2017beyond}. Studies of proactive smart speakers similarly find that the desirability of an intervention varies with its timing, usefulness, ongoing activity, and social context, as well as with individual differences in how much initiative users welcome~\cite{cha2020hello,wei2021understanding,zargham2022circumstances}. Beyond deciding whether and when to intervene, the manner of interaction also shapes acceptance. Work on conversational agents shows that language style and level of detail affect perceived competence, comfort, and social expectations~\cite{clark_what_2019,chin2024like,nass1994computers}, while proactive systems can better preserve user agency by asking, confirming, or explaining before taking action~\cite{oh2024better,liu_understanding_2023}. Long-term studies further show that assistant use and expectations develop across daily routines, making fixed or population-wide design rules insufficient for sustained interaction~\cite{bentley2018understanding}. These findings draw from foundational theories of mixed-initiative interaction~\cite{horvitz1999principles} and social presence~\cite{nass1994computers}, which suggest that agents must navigate the ``social complexity'' of human environments to avoid being perceived as intrusive. Clear guidance now exists regarding interaction timing, communication style, and user control~\cite{liu_understanding_2023, yang2024talk2care, bentley2018understanding}. However, most existing systems implement these principles in isolation, often treating autonomy and transparency as static tradeoffs. This fragmented approach fails to capture how these dimensions interact and vary across individuals during daily usage. Our work addresses this gap by modeling user preferences as structured, multi-dimensional signals that jointly influence proactive decisions.

\subsection{LLM-Driven User Behavior Simulation }
Studying preference alignment at scale through real-world deployments is challenging due to privacy concerns and deployment costs. Recent work addresses this by using large language models to simulate human participants in high-fidelity virtual environments~\cite{park2023generative, park2024generative}. By maintaining persona profiles, memories, daily plans, and evolving interaction contexts, these generative agents can produce controlled and repeatable behavioral traces over longer time periods than are typically feasible in human-subject studies. Large-scale platforms such as OASIS further demonstrate that LLM agents can be situated within dynamic social environments to study population-level interaction patterns and emergent group behavior under controlled conditions~\cite{yang2024oasisopenagentsocial}.
This capability is anchored in the concept of algorithmic fidelity~\cite{argyle2023out}, which posits that LLMs can mirror complex human demographic and psychological distributions to produce authentic interaction traces. Studies of persona-conditioned simulation show that persona information can improve correspondence with human judgments and support structured value and reasoning profiles, including moral frameworks that generalize to unseen decision scenarios, although the resulting fidelity depends on how well the selected persona attributes capture actual human variation~\cite{hu2024quantifying, An2025MoralReason}. However, simulated behavior remains sensitive to the underlying model, persona construction, prompting strategy, environment design, and validation process.
For proactive assistants, LLM-driven simulators like GIDEA~\cite{gidea_imwut, gidea_sensys} generate ``silicon samples,'' allowing researchers to explore the vast space of user-assistant interactions that would be logistically impossible to observe in the real world~\cite{lu2024proactive}. {More recently, ProPerSim simulates persona-conditioned daily activities and repeated assistant interventions, using longitudinal feedback to improve the timing and personalization of proactive recommendations~\cite{kim2025propersim}. By prioritizing preference robustness over physical robustness, these simulations provide a safe, scalable sandbox for testing how proactive assistants align with diverse human values before reaching a physical device.

\subsection{Training Context-Aware and Preference-Aligned Assistants} 
Recent systems demonstrate effective proactive assistance through two primary approaches. First, context-aware systems use multimodal sensing and historical information to determine when assistance is needed and what action should be taken. Platforms such as SocialMind~\cite{yang2025socialmind} and ProAgent~\cite{yang2025proagent} use wearable hardware to capture rich contextual cues. While effective, these require instrumented environments and lack the scalability of mobile-first solutions. 
Second, post-training alignment techniques such as Reinforcement Learning from Human Feedback (RLHF) and Direct Preference Optimization (DPO) refine behavior using preference signals~\cite{christiano2017deep, rafailov2023direct, ouyang2022training}. 
Proactive Agent applies this data-driven approach through ProactiveBench, using human accept--reject judgments to train a reward model and fine-tune whether an agent should offer assistance~\cite{lu2024proactive}; ContextAgentBench further evaluates proactive prediction and tool use from multimodal daily-life context~\cite{yang2025contextagent}. However, standard alignment usually aggregates feedback across annotators, producing behavior aligned with population-level averages rather than potentially conflicting individual preferences. Personalized alignment methods address this limitation through lightweight user models, decomposed preference dimensions, retrieval from user histories, or user-specific parameter-efficient modules~\cite{li2024personalized, jang2023personalized, salemi2024lamp, tan2024democratizing}. These approaches improve individual alignment, but often require sufficient user histories, explicit preference annotations, or separate optimization for each user, leaving both cold-start and resource-constrained deployment challenges.
Inference-time activation interventions offer a lighter alternative by modifying a frozen model's internal representations without full parameter updates~\cite{li2023inference, turner2024activation}. Prior work, however, primarily steers fixed global attributes such as truthfulness, sentiment, or personality rather than multiple user-specific preferences that evolve through interaction.
Our work addresses this gap through a simulation-first training strategy: LLM-driven simulation supplies large-scale, structured preference interactions for population-level learning, followed by lightweight on-device activation steering from individual feedback. This population-to-individual design combines a strong cold start with efficient, privacy-preserving adaptation without maintaining a separately trained model for every user.

\begin{table*}[h]
\centering
\footnotesize
\caption{Five selected preference categories that shape proactive assistant behavior.}
\vspace{-10pt}
\label{tab:preference_categories}
\setlength{\tabcolsep}{3pt}
\begin{tabular}{p{0.18\textwidth}  p{0.42\textwidth} p{0.38\textwidth}}
\toprule
\textbf{Category} & \textbf{Rationale} & \textbf{Example} \\
\midrule
Scheduling Preference & Temporal rules for \textbf{when} interactions are welcome & ``Notify about deliveries after 6pm on weekdays'' \\[0.5ex]
Domain Prioritization & \textbf{What} user values across task domains & ``Prioritize energy efficiency over convenience'' \\[0.5ex]
Autonomy Level & \textbf{How much} control user retains over actions & ``Always ask before purchases over \$50'' \\[0.5ex]
Communication Style & \textbf{How} responses are phrased in tone and detail & ``Keep responses brief, skip explanations'' \\[0.5ex]
Context Adaptation & Adaptation to \textbf{where} user is across environments & ``Different lighting for bedroom vs. living room'' \\
\bottomrule
\end{tabular}
\end{table*}

\section{Motivation and Problem Formulation}
\label{sec:motivation_problem_formulation}
\subsection{Designing Proactive Smart Assistants}
\label{sec:designing proactive}
\textcolor{myblue}{To design proactive smart assistants,} we model proactivity as a sequence of intervention opportunities that arise during daily activities. At each opportunity, the assistant makes two coupled decisions: (1) whether to initiate a proactive interaction or remain silent, and (2) if it initiates, how to respond (for example content, level of detail, and tone) and how much initiative to take. Even when the underlying task content is reasonable, users may react negatively if the assistant intervenes at a bad time, acts without enough user control, or uses an unwanted communication style. Repeated mismatches reduce trust and can lead users to ignore suggestions or disable proactive features.

\textcolor{myblue}{Prior human-participant studies of proactive assistants, mainly spanning Wizard-of-Oz and storyboard-based experiments, suggest that users judge proactive behavior holistically rather than along a single factor. To reflect this structure,}  we organize user preferences into five recurring categories \textcolor{myblue}{that capture common dimensions of proactive assistant behavior.} 
\textcolor{myblue}{
\begin{itemize}
    \item \textbf{Scheduling preference} captures when interventions are welcome, and is consistently identified as a primary factor in whether proactive behavior is accepted or rejected~\cite{cha2020hello, zargham2022circumstances, bentley2018understanding}.
    \item \textbf{Domain prioritization} describes what the user values across task domains, since users differ substantially in which goals or task areas they want an assistant to track and surface~\cite{wei2021understanding, SCHIAFFINO2004129}.
    \item \textbf{Autonomy level} reflects how much control the user retains over assistant actions, a recurring tension in proactive design where users often prefer confirmation before action, but differ in how much initiative they want the assistant to take~\cite{oh2024better, zargham_i_2022, liu_understanding_2023}.
    \item \textbf{Communication style} concerns how responses are phrased in tone and detail, which can influence perceived intelligence, trust, and comfort independently of the task content~\cite{chin2024like, clark_what_2019, bentley2018understanding}.
    \item \textbf{Context adaptation} captures how proactive behavior should change across environments, times, and ongoing activities, reflecting the expectation that assistants should adjust to situations rather than apply uniform policies everywhere~\cite{cha2020hello, wei2021understanding, zargham_i_2022}.
\end{itemize}
These categories often co-occur within the same interaction. For example, a user may want brief reminders (\emph{communication style}) but only at natural breaks (\emph{scheduling}), while still expecting the assistant to ask before taking actions (\emph{autonomy}). Table~\ref{tab:preference_categories} summarizes the categories and representative examples.
}

\begin{table*}[h]
\footnotesize
\centering
\caption{\textcolor{myblue}{Comparison of existing dataset resources for proactive preference learning. We compare whether each resource provides diverse user profiles, longitudinal preference continuity, explicit preference schema, and grounded daily activity context.}}
\label{tab:dataset_limitations}
\renewcommand{\arraystretch}{1.08}
\setlength{\tabcolsep}{5pt}
\begin{tabularx}{\textwidth}{>{\raggedright\arraybackslash}p{0.3\textwidth} c c c c}
\toprule
\multicolumn{1}{c}{\small {\textbf{Dataset}}} &
\textbf{\makecell{\textcolor{myblue}{Diverse}\\\textcolor{myblue}{User Profiles}}} &
\textbf{\makecell{\textcolor{myblue}{Longitudinal}\\\textcolor{myblue}{Preference Traces}}} &
\textbf{\makecell{\textcolor{myblue}{Explicit}\\\textcolor{myblue}{Preference Schema}}} &
\textbf{\makecell{\textcolor{myblue}{Grounded Daily}\\\textcolor{myblue}{Activity Context}}} \\
\midrule
\textbf{Activity-Centered:} CASAS~\cite{cook2012casas}, Home-Assistant-Requests~\cite{acon96_home_assistant_requests}
& \textcolor{red}{\ding{55}}
& \textcolor{red}{\ding{55}}
& \textcolor{red}{\ding{55}}
& \textcolor{ForestGreen}{\checkmark} \\
\midrule
\textbf{Preference-Centered:} PrefEval~\cite{zhao2025llms}, 
LaMP~\cite{salemi2024lamp}, PersonaLens~\cite{zhao2025personalens}
& \textcolor{ForestGreen}{\checkmark}
& \textcolor{red}{\ding{55}}
& \textcolor{ForestGreen}{\checkmark}
& \textcolor{red}{\ding{55}} \\
\midrule
\textbf{Conversational:} PersonaChat~\cite{zhang-etal-2018-personalizing}, MultiWOZ~\cite{budzianowski2020multiwozlargescalemultidomain}, RoleLLM~\cite{wang2024rolellm}
& \textcolor{ForestGreen}{\checkmark}
& \textcolor{red}{\ding{55}}
& \textcolor{red}{\ding{55}}
& \textcolor{red}{\ding{55}} \\
\midrule
\textbf{Ours (Simulation-based Dataset)}
& \textcolor{ForestGreen}{\checkmark}
& \textcolor{ForestGreen}{\checkmark}
& \textcolor{ForestGreen}{\checkmark}
& \textcolor{ForestGreen}{\checkmark} \\
\bottomrule
\end{tabularx}
\end{table*}

\subsection{Challenges for Proactive Preference Learning}
\label{sec:challenges}
\textcolor{myblue}{The multi-dimensional nature of proactive preferences creates two interrelated challenges for building real assistants. Although prior HCI studies have identified important factors that shape proactive assistant behavior, most existing work remains at the population level. Wizard-of-Oz studies, storyboard interviews, and controlled evaluations are effective for revealing broad trends in how users respond to timing, autonomy, communication style, and contextual variation, but they do not directly solve the cold-start problem faced by real assistants in deployment. In practice, assistants must begin interacting with individual users before sufficient personal history is available, even though early proactive mistakes can be especially costly for trust and long-term adoption. This creates a gap between population-level findings about good proactive behavior and the need for user-specific adaptation in real use. To address this gap, we adopt a two-stage learning strategy that first captures shared preference structure at the population level and then supports lightweight adaptation to individual users during deployment.}

\textcolor{myblue}{A second challenge is the absence of suitable training data. To support proactive preference learning, such a dataset should satisfy four properties: (1) \textbf{diverse user profiles}, so that preference expression varies across backgrounds and routines; (2) \textbf{longitudinal preference traces}, so that preferences can be observed across repeated daily situations rather than only in isolated one-shot interactions; (3) \textbf{explicit preference schema}, so that preference signals can be organized into learnable categories; and (4) \textbf{grounded daily activity context}, so that proactive decisions remain tied to what the user is doing at the time of intervention. Existing resources do not satisfy these requirements simultaneously (Table~\ref{tab:dataset_limitations}), which motivates our dataset construction. Rather than replacing real user data, this constructed dataset is designed to support the population-level learning stage of our framework by providing a structured training signal before later adaptation to individual users.}

\subsection{Motivation: Simulation as a Training Ground for Proactive Assistants}
\label{sec:motivation_simulation}
\textcolor{myblue}{These challenges motivate our central hypothesis: simulation can serve as a practical training ground for proactive assistants if it is used to construct activity-grounded interaction traces that expose when and how preferences are expressed.} Simulation has long been used as a practical tool for training and testing intelligent systems before real-world deployment. Simulated environments are routinely used to train control policies, explore failure cases, and accelerate learning while avoiding the cost and risk of physical experimentation. A similar opportunity exists for proactive smart assistants, where collecting large-scale, longitudinal interaction data from real users is costly, privacy-sensitive, and difficult to scale. In particular, proactive behavior is  \textcolor{myblue}{difficult to test early in deployment because poorly timed or intrusive actions can quickly erode trust, limiting both data collection and iterative refinement exactly when these systems most need user feedback.}

Recent advances in LLMs enable a complementary form of simulation focused on human behavior rather than physical dynamics. Platforms such as GIDEA demonstrate that LLM-based generative agents can simulate human participants in assistant studies with high semantic fidelity~\cite{gidea_imwut, gidea_sensys}. \textcolor{myblue}{In particular,} GIDEA reproduces findings from \textcolor{myblue}{published HCI studies on smart assistant interaction, generating traces that align with human feedback across diverse routines and multi-session scenarios.}
\textcolor{myblue}{These results suggest that LLM-driven simulation can capture meaningful patterns in how users react to proactive assistant behavior.} 
\textcolor{myblue}{Building on this, we use simulation to construct activity-grounded interaction traces that expose when and how preferences are expressed across a large and diverse synthetic population. This simulation-first approach provides the structured training signal needed to address both challenges identified above: it supplies population-level preference structure for a strong cold start, and it produces data that satisfies the four requirements outlined in Section~\ref{sec:challenges} without collecting real user logs.}

\subsection{Problem Formulation}
\label{sec:problem formulation}
We consider proactive assistants that run on personal devices and support a single user across daily activities. At each potential intervention point \(t\), the assistant observes the current activity context \(a_t\), recent interaction history \(h_t\), and situational signals such as time or location. The assistant must decide whether to remain silent or initiate a proactive action and, if it initiates, how to respond so that the user experiences the interaction as appropriate and helpful.

The preferences that guide these decisions are latent. They differ across users, depend on context, and may change as the user interacts with the assistant. Guided by prior HCI studies and our analysis of simulated interaction traces, we represent preferences using the five recurring dimensions in Table~\ref{tab:preference_categories}. We denote the set of preference dimensions as
\[
C = \{C_{\text{scheduling}}, C_{\text{domain}}, C_{\text{autonomy}}, C_{\text{comm}}, C_{\text{context}}\}.
\]
These dimensions recur across domains and user populations, but individuals differ in how important each dimension is and in what they want within each dimension.

For each user $u_i$, we represent their long-term tendencies across these dimensions by a vector $P_i \in \mathbb{R}^{|C|}$, which encodes how strongly the user typically weights each dimension and how preferences are expressed in practice. While the dimensions in $C$ are shared at the population level, the values of $P_i$ differ across users and may change over time as interaction feedback accumulates.
During a specific interaction in activity context $a_j$, only a subset of dimensions in $C$ may be relevant. The assistant therefore faces two coupled inference problems: identifying which preference dimensions are active in the current situation, and interpreting user-specific preferences within those dimensions well enough to decide whether and how to intervene. Errors in either step can lead to poorly timed, overly intrusive, or mismatched proactive behavior.

\textcolor{myblue}{We formulate proactive assistant personalization as a population-to-individual learning problem with multidimensional preference structure and on-device deployment constraints. This leads to a two-stage design in which category-structured supervision is used to learn shared preference representations at the population level, while category-specific activation steering supports lightweight individual adaptation without per-user retraining.}
In Stage~1, the assistant learns shared preference structure from large-scale synthetic interactions generated by LLM-simulated users, thereby providing a strong cold start without relying on real user logs.
In Stage~2, the assistant adapts to each user on device by adjusting how it attends to the preference dimensions based on lightweight interaction feedback. The following sections describe how we construct the synthetic preference dataset and implement this two-stage personalization pipeline under mobile and privacy constraints.


\begin{figure*}[h]
    \centering
    \includegraphics[width=\linewidth]{figures/overview.pdf}
    \vspace{-15pt}
    \caption{System Overview}
    \label{fig:overview}
    \vspace{-15pt}
\end{figure*}

\section{Overview}


Figure~\ref{fig:overview} presents the full system design. We begin by generating a large-scale synthetic interaction dataset (\S~\ref{sec:synthetic_dataset_generation}) that captures generalizable patterns in five preference categories across $1{,}000$ personas, providing the population-level foundation required for proactive assistant behavior. Building on this dataset, we introduce a two-stage personalization framework (\S~\ref{sec:two_stage_personalization_framework}): Stage~1 learns category-level preference understanding through supervised fine-tuning (\S~\ref{sec:stage1}), and Stage~2 adapts behavior for each individual through lightweight activation vector steering driven by real interaction feedback (\S~\ref{sec:stage2}). We then deploy this framework on edge devices (\S~\ref{sec:implementation}), enabling efficient, privacy-preserving personalization under mobile hardware constraints. The system is evaluated through simulation-based experiments (\S~\ref{sec:evaluation}) to assess preference understanding and proactive decision quality. Finally, a two-part user study (\S~\ref{sec:user_study}) validates the real-world applicability of our approach, examining how users perceive personalized behaviors and how adaptation influences satisfaction during repeated interactions.
In the following sections, we introduce each component in detail.

\section{Synthetic Interaction Dataset Generation}
\label{sec:synthetic_dataset_generation}

\textcolor{myblue}{We organize the synthetic data pipeline into three steps: persona pool construction, preference trace simulation, and simulated data validation. This structure reflects the role of each component: grounded persona profiles provide structured background diversity, GIDEA serves as the interaction simulation substrate, and the resulting synthetic traces are evaluated for population-level plausibility.}

\subsection{Persona Pool Construction}
\label{sec:persona construction}
We begin by constructing a persona pool to ground interaction behavior at the population level. The pool is derived from the Nemotron-Personas-USA v1.0 dataset~\cite{nemotron_dataset}\footnote{Nemotron-Personas-USA v1.1 was released on October 28, 2025 and expanded the dataset to approximately 1 million personas. In this work, we use the v1.0 snapshot for persona pool construction and sampling to ensure a stable and well-defined source pool.}, which is constructed from U.S. Census Bureau statistics using probabilistic graphical models.

We sample 1,000 personas for this study, and our sampled group excludes under-18 personas. The dataset provides each persona with a coherent set of demographic and lifestyle attributes, such as age range, occupation category, education level, and geographic region. \textcolor{myblue}{These attributes are not treated as ground-truth user preferences.}
\textcolor{myblue}{Instead, they provide structured background conditioning that shapes daily routines, constraints, and language use during simulation. For example, different attributes can lead to distinct recurring activities and opportunities for proactive support, such as pet-related routines for a user who owns a dog versus one who does not. Thus, the persona source introduces structured diversity in everyday contexts, while preference expressions are elicited later through the interaction simulation process.}

\subsection{Preference Trace Simulation with GIDEA}
\label{sec:preference trace}
\noindent \textcolor{myblue}{\textbf{Simulation Substrate:}
GIDEA provides the agent-based interaction environment in which grounded personas, daily activity schedules, and proactive assistant proposals are instantiated at scale. Building on this, we introduce a structured elicitation process that converts persona-conditioned activity interactions into explicit preference supervision for proactive assistant learning, organized around the five categories in Table~\ref{tab:preference_categories}.}

\noindent \textbf{User Activity Construction:}
For each persona, we generate a week-long activity schedule using GPT-4.1. \textcolor{myblue}{We use eight activity types as a coverage and annotation schema rather than as fixed sampling quotas: productivity, health, cooking, entertainment, transport, cleaning, social, and miscellaneous.
These categories are drawn from real-world studies of proactive assistant interaction contexts~\cite{cha2020hello} and are intended to ensure that generated routines span both common daily activities and situations in which proactive support is likely to arise. During generation, the model is conditioned on the persona profile and asked to construct a coherent weekly routine. Activities are then assigned to one of the eight categories based on their semantic content. Thus, the final distribution emerges from persona-conditioned schedule generation rather than being forced to match a predefined population-level time-use distribution. The resulting schedules provide temporal structure for identifying potential intervention points, while natural variation across days helps ensure that preference expressions are not tied to a single recurring situation. We analyze the data quality in Section~\ref{sec:data_quality}.}

\noindent \textcolor{myblue}{\textbf{Persona-Guided Preference Elicitation:}}
\textcolor{myblue}{For each persona and activity schedule, we identify potential proactive intervention points throughout the week. At each point, the assistant generates a candidate proactive action based on the current activity context. The simulated user then reacts to this proposal from the persona’s perspective by evaluating whether the behavior is appropriate, explaining the reasoning, and describing how the interaction should be delivered if intervention is welcome.}
\textcolor{myblue}{This process elicits preference signals as feedback on concrete assistant behavior rather than as free-form dialogue. The resulting feedback is organized using the five preference categories introduced in Table~\ref{tab:preference_categories}. For each intervention point, the simulation produces three coupled outputs: (1) the subset of active preference categories, (2) a natural-language description capturing the user’s reasoning, and (3) a preferred assistant response aligned with the persona’s needs and activity context. }

\textcolor{myblue}{In this way, the assistant proposal defines a concrete interaction target, and the simulated user feedback transforms it into structured preference supervision. Across 1,000 personas and week-long schedules, this process yields approximately 100,000 training examples, each consisting of a persona profile, activity context, preference categories, preference description, and preferred assistant response. An end-to-end example is shown in Appendix~\ref{appendix:example}.}

\noindent \textcolor{myblue}{\textbf{Cross-Session Continuity:}}
\label{sec:cross-session continuity}
 \textcolor{myblue}{To support cross-session consistency,} each persona maintains an episodic memory using Reflexion-based reasoning~\cite{shinn2023reflexionlanguageagentsverbal}. After a block of interactions, the persona summarizes which assistant behaviors were helpful or disruptive, \textcolor{myblue}{and these summaries are incorporated into later prompts. We use this mechanism because our goal is not to generate isolated one-shot preferences, but to simulate repeated interactions in which earlier assistant behavior can shape later reactions. A simpler setup without episodic memory would treat each interaction more independently and would make it harder to preserve preference traces across recurring daily situations. In this work, we therefore use episodic memory as a lightweight way to maintain contextual continuity across sessions, rather than as a claim of full human-like memory.}

This representation is used directly in Stage~1 to train category prediction and response generation, and it provides the population-level signals that Stage~2 adapts to individual users during on-device personalization.


\begin{figure}[h]
    \centering\begin{minipage}[t]{0.49\textwidth}
        \vspace{0pt}
        \centering
        \scriptsize
        \captionof{table}{Distributional similarity between the sampled 1,000 personas and the 100k adult source snapshot. Max $|\Delta|$ reports the largest category-level difference in percentage points.}
        \label{tab:persona_sampling_summary}
        \setlength{\tabcolsep}{3pt}
        \renewcommand{\arraystretch}{1.03}
        \resizebox{\linewidth}{!}{%
        \begin{tabular}{lcccc}
        \toprule
        \textbf{Attribute} & \textbf{\# Groups} & \textbf{JS} & \textbf{TV} & \textbf{Max $|\Delta|$ (pp)} \\
        \midrule
        Age group          & 6  & 0.032 & 0.036 & 1.80 \\
        Gender             & 2  & 0.014 & 0.017 & 1.67 \\
        Education          & 7  & 0.048 & 0.037 & 2.29 \\
        Marital status     & 5  & 0.020 & 0.014 & 1.43 \\
        Occupation group   & 8  & 0.028 & 0.025 & 1.49 \\
        Region             & 5  & 0.037 & 0.034 & 2.97 \\
        Top states + Other & 11 & 0.046 & 0.035 & 1.70 \\
        \bottomrule
        \end{tabular}
        }

        \vspace{2pt}
        \begin{minipage}{0.95\linewidth}
        \scriptsize
        \emph{Note.} JS denotes Jensen--Shannon distance, and TV denotes total variation distance. Percentages are computed within each dataset.
        \end{minipage}
    \end{minipage}
    \hspace{0.03\textwidth}
    \begin{minipage}[t]{0.46\textwidth}
        \vspace{0pt}
        \centering
        \includegraphics[width=1\linewidth,height=0.16\textheight]{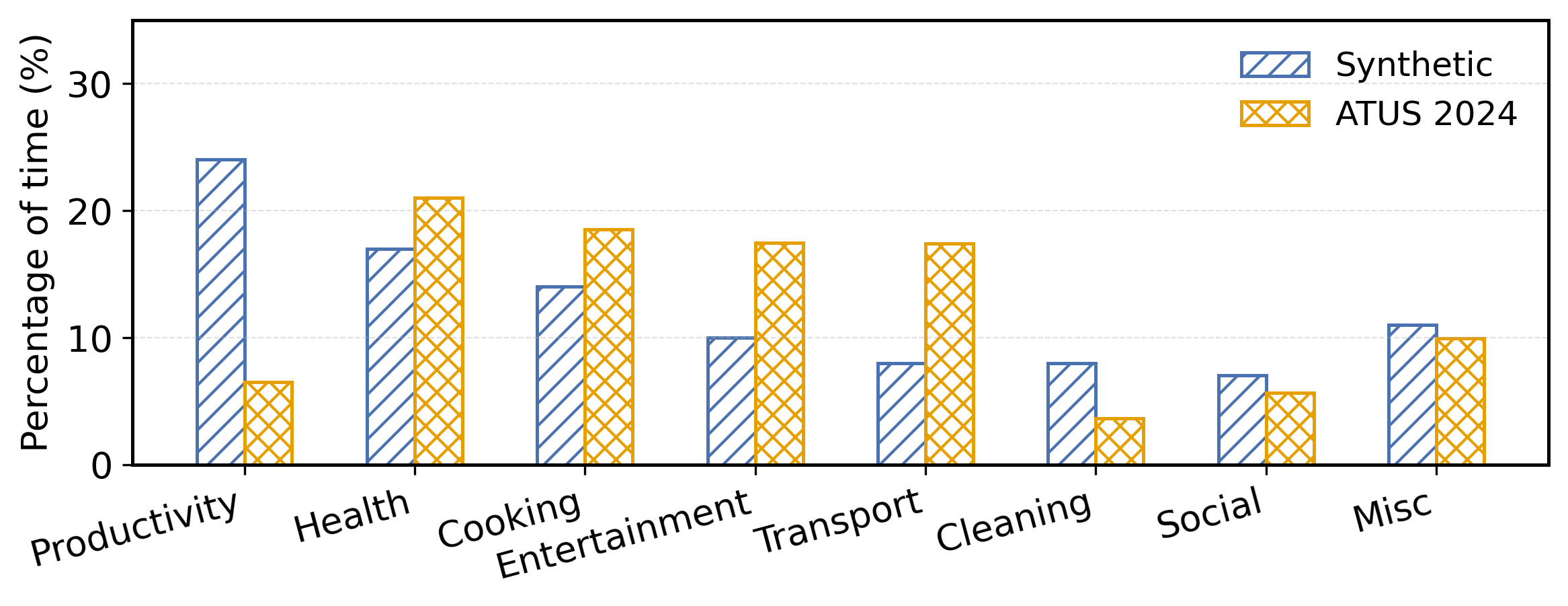}
        \captionof{figure}{Activity-category distribution of synthetic schedules compared with ATUS 2024 after reweighting ATUS respondents to match the sampled persona cohort.}
        \label{fig:activity}
    \end{minipage}
\end{figure}

\subsection{Dataset Quality and Validation}
\label{sec:data_quality}
\textcolor{myblue}{We evaluate the synthetic dataset with a focus on population-level plausibility rather than full individual-level behavioral realism. Accordingly, the validation below examines three aspects of the generated dataset: persona sampling validity, activity distribution realism, and activity-grounded preference plausibility. These checks assess whether the dataset reflects plausible aggregate activity structure, broad persona diversity, and coherent activity-conditioned preference traces, while leaving deeper validation of persona-specific behavioral realism to future work.}

\noindent\textbf{Persona Sampling Validity:}
\textcolor{myblue}{We compare the sampled 1,000 personas against the exact 100k adult source snapshot of Nemotron-Personas-USA used for sampling. Specifically, we compare marginal distributions across age group, gender, education, marital status, occupation group, region, and top states. Table~\ref{tab:persona_sampling_summary} summarizes distributional similarity between the sampled personas and the source snapshot, while Appendix~\ref{app:persona_sampling_validity_details} Table~\ref{tab:persona_sampling_detailed} reports the full category-level comparison. Across these attributes, the sampled group closely tracks the source pool. All observed category-level differences are small, with the largest absolute deviation below 3 percentage points. Distribution-level distances are also low: Jensen--Shannon distances range from 0.014 for gender to 0.048 for education, and total variation distances remain below 0.04 across all reported attributes. These results indicate that the sampled personas preserve the main demographic structure of the eligible adult source pool.}

\textcolor{myblue}{We further assess multivariate coverage using a UMAP projection of joint persona representations derived from concatenated persona text fields. As shown in Appendix~\ref{app:persona_sampling_validity_details} Figure~\ref{fig:persona_umap}, the sampled personas appear across both dense and sparse regions of the source pool, suggesting broad coverage of the source persona space beyond individual marginal attributes. This analysis does not establish individual-level behavioral realism; rather, it provides a qualitative coverage check showing that the simulated population broadly spans the adult persona source pool before interaction traces are generated.}

\noindent\textbf{Activity Distribution Realism:}
\textcolor{myblue}{Figure~\ref{fig:activity} compares the generated activity distribution against the 2024 American Time Use Survey (ATUS), after mapping both sources to the same eight broad activity categories. To make the comparison appropriate for our simulated cohort, we reweight ATUS respondents to align with the sampled 1,000 personas on overlapping observable attributes, including age group, gender, education, spouse-status proxy, and workforce status. This yields a persona-matched ATUS reference group rather than a comparison to the national ATUS sample in the aggregate.}

\textcolor{myblue}{Under this aligned benchmark, the synthetic schedules cover all eight activity categories and preserve a broad daily-life structure, while showing systematic shifts toward assistant-relevant contexts. In particular, work and productivity is overrepresented (24.0\% synthetic vs.\ 6.5\% persona-matched ATUS), as is cleaning and maintenance (8.0\% vs.\ 3.6\%). Other categories remain well represented, including health and wellness (17.0\% vs.\ 21.0\%), cooking and kitchen (14.0\% vs.\ 18.5\%), entertainment and media (10.0\% vs.\ 17.4\%), transportation and commute (8.0\% vs.\ 17.4\%), social and communication (7.0\% vs.\ 5.7\%), and miscellaneous activities (11.0\% vs.\ 9.9\%). We therefore treat the synthetic schedules as population-level plausible for model training.}


\noindent\textbf{Activity-Grounded Preference Plausibility:}
\textcolor{myblue}{To complement the distribution-level checks above, we conducted a small-scale human sanity check to assess whether generated preferences remain plausible when conditioned on both persona profiles and activity contexts. Five raters reviewed 600 unique persona-activity-preference examples, producing 1,000 total ratings. Each example included a persona profile, an activity/context snippet, and the generated preference statement. Raters first judged each example as usable, borderline, or unusable for the intended preference-learning task, and then scored profile-preference alignment, context relevance, specificity, internal consistency, realism, and diversity on five-point scales.}

\textcolor{myblue}{Across all ratings, 67.4\% of examples were marked usable, 24.9\% borderline, and 7.7\% unusable; thus, 92.3\% were judged at least acceptable. The strongest ratings were for context relevance (M=4.07/5), internal consistency (M=4.03/5), realism (M=4.00/5), and profile-preference alignment (M=3.88/5), suggesting that most generated preferences were coherent and appropriate for the provided activity context. In the 100-example overlap subset rated by all five raters, 99\% of examples were judged acceptable by a majority of raters, while only 1\% received a majority unusable judgment. We treat this check as evidence of activity-grounded plausibility rather than proof of full individual-level behavioral realism.}

\section{Two-Stage Personalization Framework}
\label{sec:two_stage_personalization_framework}
\subsection{Motivation for a Two-Stage Approach}

Proactive assistants on edge devices must be useful from the first interaction and adapt quickly to user preferences, but they also face tight compute and privacy limits. Large models can encode detailed preferences but are too heavy for on-device use, whereas small models fit locally but see few interactions per user and cannot be retrained often. Analysis of our synthetic dataset with 1,000 personas reveals population-level patterns in proactive preferences. Across personas, preferences related to scheduling, domain prioritization, autonomy level, communication style and context adaptation follow similar linguistic and behavioral patterns. Individual users primarily differ in which of these categories are salient in a given activity and in the direction or strength of each preference. This observation motivates a decomposition into shared category-level representations and user-specific weighting over these categories.

A single-stage approach that directly fine-tunes an assistant on user feedback using reinforcement learning from human feedback (RLHF)~\cite{ouyang2022training} ignores this structure and is difficult to apply on edge devices. Such methods require centralized interaction logs, maintain per-user model variants, and provide a weak cold start. In our experiments (\S~\ref{sec:overall_performance}), direct response fine-tuning mainly improves surface fluency but lags behind structure-aware training in preference understanding and temporal appropriateness.
We therefore adopt a two-stage design. Stage~1 trains a compact 3B model on large-scale synthetic data to map from persona and activity context to active preference categories and preference-aware responses. This training is performed once offline to obtain a single population-level model with reasonable default behavior. Stage~2 keeps the backbone frozen and adapts to each user through lightweight on-device activation steering that adjusts how strongly the model attends to each preference category based on simple interaction signals, without changing base parameters. This separation provides a strong cold start, avoids sending user logs to a server, and enables fast, reversible personalization under mobile hardware constraints.

\subsection{Stage 1: Population-Level Learning}
\label{sec:stage1}
\textcolor{myblue}{Unlike standard response-only SFT, Stage~1 does not train the model solely to imitate assistant outputs. Instead, it uses category-structured supervision so that the model learns shared representations of recurring preference dimensions across users and activities. This structure is important because proactive assistant behavior depends not only on response quality, but also on which preference dimensions are active in a given situation.}

\noindent \textbf{Inputs.}
Each training example consists of:
(i) a persona profile describing demographic and background attributes,
(ii) an activity context specifying the current situation in which a proactive intervention may occur,
(iii) a set of active preference dimensions $C_k \in C$, and
(iv) a natural-language assistant response that reflects how the assistant should act under these preferences.
All inputs are derived from the synthetic dataset described in \S~\ref{sec:synthetic_dataset_generation}.

\noindent \textbf{Training Objective.}
We train a single language model using supervised fine-tuning with a multi-task objective. The model is optimized to (1) generate an appropriate proactive response and (2) recognize which preference dimensions are relevant in the given context. Preference categories serve as auxiliary supervision during training, encouraging the model to attend to distinct dimensions such as timing, autonomy, and communication style when encoding the interaction context. Importantly, these categories do not form an explicit decision pipeline at inference time; instead, they shape shared internal representations that directly condition response generation.

\noindent \textbf{Model and Optimization.}
We fine-tune the Llama-3.2-3B-Instruct backbone using supervised fine-tuning with Low-Rank Adaptation (LoRA), keeping the base model weights frozen to meet on-device memory constraints. Training proceeds in two phases. In the first phase, the model is exposed to individual category–preference pairs, allowing it to learn how each preference dimension is expressed in language across personas and activities. In the second phase, the model is trained on full examples that combine multiple active categories with complete proactive responses, refining its ability to integrate preference signals into coherent assistant behavior.

\noindent \textbf{Outputs.}
The output of Stage~1 is a single population-level assistant model represented by a set of LoRA adapter weights. These adapters capture shared preference structures across users and provide reasonable default behavior for new users and unseen activity contexts. The resulting model can be deployed directly on edge devices and serves as the fixed backbone for individual adaptation in Stage~2.

\noindent \textbf{Role in the Overall Framework.}
Stage~1 establishes a strong cold start by teaching the model how preferences are structured and expressed at the population level. It does not encode user-specific priorities. Instead, it produces representations that are explicitly organized around preference dimensions, enabling efficient and interpretable individual adaptation through activation-based steering in Stage~2.

\subsection{Stage 2: Individual Adaptation}
\label{sec:stage2}

\textcolor{myblue}{Existing activation-steering methods typically target a single global attribute using fixed directions learned offline. In contrast, our setting requires online adaptation across multiple preference dimensions that jointly shape proactive behavior. We therefore extend steering to maintain category-specific directions and strengths that are updated incrementally from natural feedback signals, enabling lightweight personalization without model retraining.}
Stage~2 adapts the population-level assistant model to individual users through activation-based steering that operates entirely on-device.

Activation steering modifies a model’s behavior by applying small, targeted perturbations to internal activations at inference time, rather than updating model parameters through gradient-based training \cite{turner2024activation, li2023inference}. Prior work shows that such interventions can reliably control behavioral attributes, such as personality expression or style, while preserving the model’s underlying language capability, and with substantially lower data and compute cost than RLHF-style optimization~\cite{zhu2025personality}.
\textcolor{myblue}{In our setting, personalization is retained over time through a compact per-user steering state rather than through parameter updates or an explicit natural-language memory module. For each user and preference category, the system maintains a steering direction and steering strength that are updated incrementally from accumulated interaction feedback and reused in future interactions. This design allows earlier feedback to continue influencing later assistant behavior while keeping the backbone model fixed.}

The goal of this stage is therefore not to relearn language, but to incrementally shift how the assistant interprets context and decides when and how to act. Existing activation-based methods typically align models along a single global attribute using fixed steering directions learned offline \cite{jiang2023personality, zhu2025personality}. In contrast, we extend activation steering to support \emph{category-specific} personalization across multiple preference dimensions and to update steering signals online from natural user feedback. Each preference category is associated with an independent steering direction and strength, enabling fine-grained, reversible adaptation of behaviors such as timing or autonomy without affecting unrelated aspects of generation.

\noindent \textbf{Inputs.}
After each proactive interaction $j$, the user may provide feedback that reflects how well the assistant’s behavior matched their expectations:

\begin{itemize}
    \item \textbf{Short response:} 
    $a_j \in \{\text{accept}, \text{reject}, \text{ignore}\}$, indicating whether the proactive suggestion was appropriate.

    \item \textbf{Category-wise satisfaction:} 
    $\mathbf{s}_j = \{ s_{j,k} \mid C_k \in C \}$, where $s_{j,k} \in [1,5]$ rates satisfaction with respect to preference dimension $C_k$ (timing, autonomy, domain prioritization, communication style, and context adaptation).

    \item \textbf{Text feedback (optional):} 
    $f_j$, a free-form natural language description of the user’s preferences or concerns when the interaction is perceived as unsatisfactory.
\end{itemize}

These signals are intentionally lightweight and optional, allowing users to correct misalignment without continuous annotation or explicit training effort.
\textcolor{myblue}{When provided, they are incorporated into the user’s persistent steering state and may continue to influence future interactions.}

\noindent \textbf{Steering Signals.}
We treat an interaction as a steering signal when the feedback indicates misalignment, either because the proactive suggestion was rejected or ignored,
$a_j \in \{\text{reject}, \text{ignore}\}$, or because dissatisfaction is expressed for at least one preference dimension:
\[
\exists\, C_k \in C \;\; \text{s.t.} \;\; s_{j,k} < \tau,
\]
where the threshold is set to $\tau = 3$ in practice. This criterion ensures that steering is triggered only when the assistant’s behavior meaningfully deviates from the user’s expectations.

\noindent \textbf{Steering Examples.}
For each steering interaction $j$ associated with a preference dimension $C_k$, we construct a contrastive text pair that captures the direction of desired change. Unlike approaches that generate full counterfactual responses, we rely directly on user feedback:

\begin{itemize}
    \item \textbf{Negative example:} 
    $x_j^{-}$, corresponding to the model-generated response $r_j^{-}$ that users rejected or ignored.

    \item \textbf{Positive example:} 
    $x_j^{+}$, constructed from the user’s textual feedback $f_j^{(k)}$, which describes how the assistant should behave for preference dimension $C_k$.
\end{itemize}
\textcolor{myblue}{The resulting contrastive pair is not used to retrain the backbone model. Instead, it is incorporated into the user’s persistent personalization state as category-specific evidence. To keep this process efficient on device, we maintain running summary statistics rather than storing all past interactions. Over time, the accumulated positive and negative examples provide a compact representation of how the user prefers the assistant to adjust its behavior along each preference dimension.}

\noindent \textbf{Objective.}
The objective of Stage~2 is to adjust the model’s internal activations so that future responses are more similar to preferred behaviors and less similar to previously rejected ones. Rather than minimizing an explicit loss function, we aim to find a direction in activation space that separates these two behaviors:
\[
\langle h_{\ell}(x_j^{+}), v_{u,\ell}^{(k)} \rangle > \langle h_{\ell}(x_j^{-}), v_{u,\ell}^{(k)} \rangle,
\]

where $h_{\ell}(\cdot)$ denotes the model activation at the selected steering layer $\ell$. Intuitively, this condition means that the internal representation of preferred behavior aligns more strongly with the steering direction than that of rejected behavior.

\noindent \textbf{Deriving Steering Directions.}
\textcolor{myblue}{To derive and update steering directions over time, we maintain running summary statistics of positive and negative examples for each user, preference dimension, and steering layer.} 
Let $h_{\ell}(x) \in \mathbb{R}^d$ denote the activation at the last token of layer $\ell$ when processing text $x$. For each layer, we compute average activations over positive and negative examples:
\[
\mu_{u,\ell}^{+,(k)} = \mathbb{E}[h_{\ell}(x_j^{+})], \qquad
\mu_{u,\ell}^{-,(k)} = \mathbb{E}[h_{\ell}(x_j^{-})].
\]
The difference $v_{u,\ell}^{(k)} = \mu_{u,\ell}^{+,(k)} - \mu_{u,\ell}^{-,(k)}$
represents how the model’s internal state must change at layer $\ell$ to move from undesired behavior toward preferred behavior. Layers with larger $\lVert v_{u,\ell}^{(k)} \rVert_2$ values exhibit stronger separability and are therefore more effective targets for steering.

\noindent \textbf{Outputs.}
For each user $u$ and preference dimension $C_k \in C$, Stage~2 maintains a small personalization state consisting of:
(i) steering directions $v_{u,\ell}^{(k)}$, and
(ii) a steering strength $\alpha_u^{(k)}$.
\textcolor{myblue}{These quantities summarize the accumulated effect of user feedback and form the persistent personalization state.}
Steering directions and strengths are updated incrementally after each steering interaction and persist across subsequent interactions for the same user.

\noindent \textcolor{myblue}{\textbf{Steering Strength Update.} For each steering interaction, the strength $\alpha_u^{(k)}$ is increased for preference dimensions that receive negative feedback, such as rejection, ignored suggestions, or satisfaction scores below the neutral threshold. Repeated feedback in the same direction increases the corresponding strength, while the value is clipped to a fixed range to prevent over-steering. When a preference dimension receives no recent supporting feedback, its strength is multiplied by a decay factor, allowing outdated signals to gradually weaken.}

\noindent \textbf{Inference-Time Steering.}
During response generation, steering is applied by modifying hidden activations at selected layers, combining contributions from all five preference dimensions:
\[
h_{\ell}' = h_{\ell} + \sum_{C_k \in C} \alpha_u^{(k)} \, v_{u,\ell}^{(k)}.
\]
This operation adjusts the model toward behaviors aligned with the user's category-specific preferences without altering the underlying language model weights.
\textcolor{myblue}{Because this steering state persists across interactions, its effect is cumulative: repeated feedback in a consistent direction strengthens its influence on future responses.}
Steering strength is bounded and decays over time, ensuring stability and allowing the system to revert to population-level behavior when feedback is sparse.
\textcolor{myblue}{In practice, decay is applied when no recent feedback supports a preference dimension, reducing the influence of outdated signals while preserving stable long-term preferences.}

\noindent \textbf{Role in the Overall Framework.}
Stage~2 converts sparse, natural user feedback into directional adjustments in activation space, while retaining this information as a persistent steering state rather than full interaction histories.
By explicitly separating inputs, outputs, and objectives, this mechanism enables fast, interpretable, and reversible personalization entirely on-device, without retraining or maintaining per-user model copies.

\begin{figure}[t] 
\centering 
\includegraphics[width=0.8\linewidth]{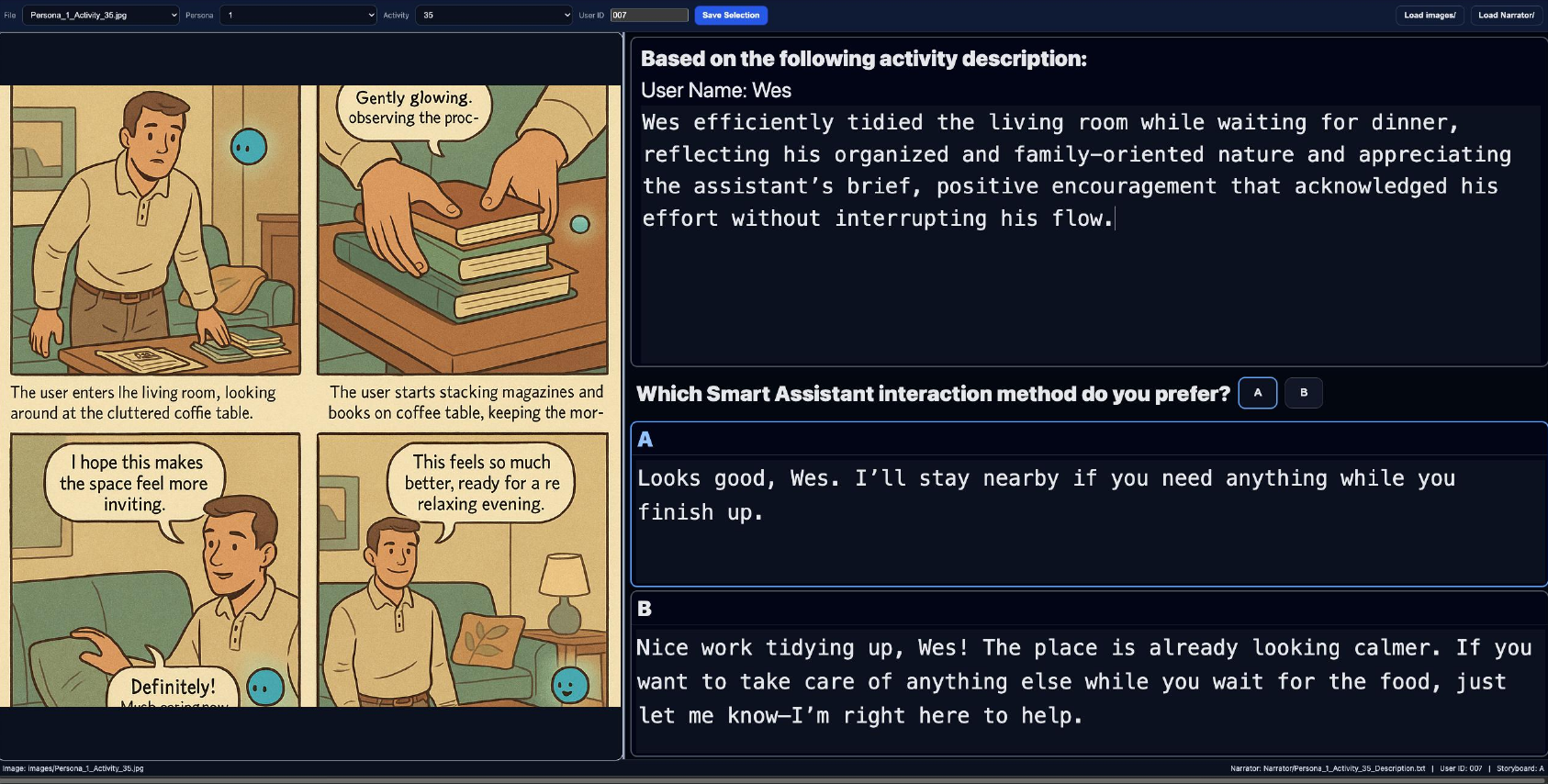} 
\vspace{-6pt} 
\caption{Customized web-based interface used in the Preference Detection Study.} 
\vspace{-16pt} 
\label{fig:study_a} 
\end{figure}

\section{Implementation}
\label{sec:implementation}

All simulation and model training are performed on a local server equipped with dual NVIDIA RTX A5000 GPUs. This environment supports synthetic persona generation, large-scale interaction simulation, and supervised fine-tuning for Stage~1 of our two-stage framework. To assess real-world feasibility, we deploy a simplified end-to-end workflow on a modern smartphone. The deployed model uses a 3B backbone with Stage~1 LoRA adapters pre-merged and quantized to 4-bit precision for on-device execution. All Stage~2 personalization is performed through activation-based steering at inference time, without modifying model weights or requiring server communication.

To support controlled data collection, the user studies are conducted using a custom web-based interface. Figure~\ref{fig:study_a} shows a representative storyboard-based task used in the studies. The interface guides participants through short visual narratives depicting everyday activities, accompanying textual descriptions, and proactive assistant interactions. 
The interaction flow mirrors the logic of the mobile prototype, ensuring that participants evaluate the same types of proactive behaviors and preference signals under controlled conditions. Additional examples of the study interface and tasks used in other study phases are provided in Appendix~\ref{sec:appendix_interface}.

\section{Evaluation}
\label{sec:evaluation}

We evaluate the proposed two-stage personalization framework using both large-scale simulation with synthetic personas and an IRB-approved human-subject study (IRB protocol 2387616-1). The evaluation has two goals: to quantify how the framework changes proactive assistant behavior under controlled conditions, and to understand how these changes are perceived by real users. We organize our study around five research questions:

\begin{rqbox}
\small
\noindent\textbf{RQ1:} How can we design metrics that capture the effectiveness of proactive human–assistant interaction?

\noindent\textbf{RQ2:} How can we design an evaluation that links simulation and real-user studies in a consistent way?

\noindent\textbf{RQ3:} To what extent does the two-stage personalization framework improve proactive assistant behavior?

\noindent\textbf{RQ4:} How do population-level training diversity and long-term interaction depth affect interaction quality, temporal appropriateness, and preference understanding?

\noindent\textbf{RQ5:} How robust is the two-stage framework across assistant model sizes and different user simulators?
\end{rqbox}

\subsection{Evaluation Metrics (RQ1)}
\label{sec:evaluation_metrics}

\textcolor{myblue}{Measuring proactive assistant behavior is challenging because success cannot be judged only by response correctness or task completion. A proactive assistant must decide \emph{when} to intervene, \emph{whether} intervention is appropriate at that moment, and \emph{how} to respond in a way that respects the user's context, autonomy, and preferences. This motivates RQ1: how can we design metrics that capture the effectiveness of proactive human--assistant interaction?}

\textcolor{myblue}{To answer this question, we evaluate effectiveness from two complementary perspectives: holistic user perception and structured behavioral alignment. First, we segment continuous human behavioral data into a sequence of candidate intervention moments, where each moment represents a fine-grained opportunity for the assistant to either intervene or remain silent. This allows us to evaluate timing decisions at the moment level rather than only at the activity level. Second, we measure the quality of delivered interactions and the assistant's ability to understand user preferences using metrics grounded in the five preference categories in Table~\ref{tab:preference_categories}.}

\textcolor{myblue}{Based on this formulation, we use one user-facing outcome metric and three structured alignment metrics. Interaction Quality Assessment (IQA) measures the user's overall perception of a delivered assistant interaction. Temporal Appropriateness Index (TAI) measures whether the assistant intervenes at acceptable moments. Categorical Alignment Score (CAS) measures whether the assistant identifies the relevant preference dimensions, and Preference Semantic Coherence (PSC) measures whether the assistant interprets the user's preference meaning correctly. Together, these metrics support method-level comparison while also explaining where improvements come from: overall perceived quality, timing alignment, category-level preference recognition, or semantic preference understanding.}

\noindent\textbf{\Circled{1} Interaction Quality Assessment (IQA).}
\textcolor{myblue}{IQA measures the user's holistic perception of a delivered proactive interaction. It is computed only when the assistant chooses to intervene and generates a response. After each delivered interaction, the user rates five aspects of the assistant's behavior on five-point Likert scales: timing appropriateness, intrusiveness, content value, contextual relevance, and respect for user autonomy. The intrusiveness item is reverse-coded so that higher values consistently indicate better interaction quality. Each item is then normalized to $[0,1]$, and IQA is computed as the average of the five normalized item scores. In this way, IQA captures how appropriate, useful, and respectful the interaction feels to the user after the assistant has acted.}

\noindent\textbf{\Circled{2} Temporal Appropriateness Index (TAI).}
\textcolor{myblue}{TAI measures whether the assistant makes the correct timing decision at each candidate intervention moment. While IQA captures a graded user rating after a delivered interaction, TAI evaluates timing as a finer-grained binary alignment problem: at each candidate moment, should the assistant intervene or remain silent? The assistant's decision is compared with the user's acceptability judgment for that moment, and TAI is defined as the fraction of candidate intervention moments where the two decisions match. For example, within the same cooking activity, an interruption while washing hands may be acceptable, whereas an interruption while cutting ingredients may be rejected. TAI captures this moment-level distinction between acceptable and unacceptable intervention timing, while IQA captures the user's softer overall judgment of the interaction once it is delivered.}

\noindent\textbf{\Circled{3} Preference Understanding (CAS \& PSC).}
\textcolor{myblue}{Preference understanding measures whether the assistant can infer the preference signals behind a user's judgment. While IQA captures perceived interaction quality and TAI captures timing alignment, CAS and PSC provide a more diagnostic view of why an interaction is or is not aligned with the user. We evaluate preference understanding through two components. The \emph{Categorical Alignment Score} (CAS) measures whether the assistant identifies which preference dimensions are active in the current context, such as scheduling preference, autonomy level, or communication style. We compute CAS as the F1 score between the predicted active preference categories and the reference categories. The \emph{Preference Semantic Coherence} (PSC) measures whether the assistant correctly interprets the meaning of the user's preference within those categories. We compute PSC as the cosine similarity between the generated preference summary and the reference preference description. Together, CAS and PSC evaluate whether the assistant understands both \emph{which} preference dimensions matter and \emph{what} the user wants within those dimensions.}

All metrics are computed at the interaction level and averaged over personas. To assess generalization, we report results separately for \emph{seen} activity contexts (present during Stage~1 training) and \emph{unseen} contexts held out during training. IQA captures overall interaction quality, while TAI, CAS, and PSC characterize how timing and preference modeling contribute to observed behavior.

We note that CAS, PSC, and TAI are intentionally aligned with the structured preference framework proposed in this paper. These metrics are therefore best interpreted as diagnostic measures that assess how well an assistant captures preference relevance, semantic interpretation, and timing decisions under the assumed preference decomposition, rather than as task-agnostic indicators of user benefit. Improvements on these metrics indicate alignment with the modeled preference structure, but do not alone establish overall interaction quality. Accordingly, we treat Interaction Quality Assessment (IQA) as the primary outcome measure. IQA reflects users’ holistic judgments of timing, intrusiveness, usefulness, and autonomy, and is collected using the same protocol in both simulation and human studies. The structured metrics are used to explain \emph{why} IQA differs across methods, rather than to replace it.

\subsection{Experimental Setup (RQ2)}

\textcolor{myblue}{The metrics above define how proactive assistant behavior is measured, but they also raise an important methodological question: where these measurements should come from. Simulation-based evaluation provides scale, repeatability, and control, but simulated users cannot fully capture real users' hesitation, attention, subjective judgment, or decision-making in proactive settings. Human studies provide direct user feedback, but they are more costly and limited in scale. We therefore use the two evaluation settings for different but connected purposes, rather than treating one as a replacement for the other. }

\noindent\textbf{Simulation-Based Assessment.}  
\textcolor{myblue}{In our evaluation, simulation serves as a large-scale methodological plausibility check rather than a replacement for human evaluation.} We simulate 1{,}000 personas sampled from our synthetic dataset, and each persona is exposed to 100 proactive interaction opportunities distributed across 10 temporal periods. \textcolor{myblue}{This controlled setup allows us to compare methods across repeated interaction trajectories, held-out activity contexts, and multiple model conditions. It helps test whether the framework behaves as intended: whether population-level learning improves cold-start behavior, whether individual adaptation changes behavior over time, and whether improvements can be explained through timing alignment and preference understanding. We interpret these simulation results as controlled evidence of method behavior, not as direct estimates of real-world user benefit.}

 Personas are instantiated using GPT-4.1 and cross-checked with Llama-3.1-70B and Claude-Sonnet-4 to reduce reliance on a single simulator and to identify major inconsistencies in simulated behavior~\cite{openai2025gpt4.1, metalLlama31_70b2024, anthropic2024claude}. The assistant under evaluation is the Llama-3.2-3B model after Stage~1 population-level fine-tuning~\cite{meta_llama3_2_3b_2024}. \textcolor{myblue}{We use this fine-tuned model to reflect a realistic deployment setting, where an assistant is expected to incorporate general preference structure before individual adaptation. Evaluating only the raw backbone would primarily test generic language capability rather than preference-aware proactive behavior.}
\textcolor{myblue}{To assess generalization, we evaluate the model using the same seen/unseen activity split described above. We also conduct a small-scale sanity check using additional compact backbones from the Llama and Qwen families to verify that the observed trends are not specific to a single base model~\cite{grattafiori2024llama, yang2025qwen3}.}

\noindent\textbf{Human User Study.}  
\textcolor{myblue}{The human study complements the simulation-based assessment by examining how real users perceive the assistant behaviors generated by the framework.} We conduct a storyboard-based user study with \textcolor{myblue}{34} participants. Participants compare personalized and non-personalized assistant responses and provide IQA ratings for each interaction. Section~\ref{sec:user_study} describes the procedure and sampling in detail; here we focus on how the study \textcolor{myblue}{complements the simulation-based evaluation}.

\noindent\textbf{Method Comparison.}  
We compare methods at two levels to separate the effects of population-level learning from individual adaptation.

\noindent\textit{Population-Level Learning:}
\begin{itemize}
    \item \textbf{Baseline:} Untuned Llama-3.2-3B.
    \item \textbf{Direct Response Training:} supervised fine-tuning to produce assistant responses without explicit preference structure.
    \item \textbf{Category-Structured Training:} supervised fine-tuning that uses our five-category preference annotations for both preference prediction and response generation.
\end{itemize}

\noindent\textit{Individual adaptation on top of the Category-Structured model:}
\begin{itemize}
    \item \textbf{+ In-Context Learning (ICL):} retrieval of the five most relevant recent interactions at inference time, without parameter updates.
    \item \textbf{+ RLHF (DPO):} parameter updates using user preferences as pairwise feedback with a DPO objective.
    \item \textbf{+ Steering:} the proposed activation-vector steering method that adjusts internal activations per user on device, without changing model weights.
\end{itemize}

\begin{table}[!t]
\centering
\caption{Performance comparison across population-level training and individual adaptation methods. CAS and PSC measure preference understanding, TAI measures timing alignment, and IQA measures overall interaction quality.}

\label{tab:comprehensive_comparison}
\scriptsize
\setlength{\tabcolsep}{3pt}
\resizebox{\columnwidth}{!}{%
\begin{tabular}{l ccc ccc cc cc}
\toprule
& \multicolumn{3}{c}{\textbf{Seen}} 
& \multicolumn{3}{c}{\textbf{Unseen}} 
& \multicolumn{2}{c}{\textbf{Avg.}} 
& \multicolumn{2}{c}{\textbf{Impr.}} \\
\cmidrule(lr){2-4} \cmidrule(lr){5-7} \cmidrule(lr){8-9} \cmidrule(lr){10-11}
\textbf{Method} 
& \textbf{CAS/PSC} & \textbf{TAI} & \textbf{IQA}
& \textbf{CAS/PSC} & \textbf{TAI} & \textbf{IQA}
& \textbf{TAI} & \textbf{IQA}
& \textbf{TAI\%} & \textbf{IQA\%} \\
\midrule
\multicolumn{11}{l}{\textit{Population-Level Baselines (No Individual Adaptation)}} \\
\midrule
Baseline (Untuned) 
& 0.295 / 0.524 & 0.156 & 0.802 
& 0.200 / 0.533 & 0.123 & 0.792 
& 0.140 & 0.797 & -- & -- \\
Direct Response 
& 0.283 / 0.596 & 0.386 & 0.865 
& 0.201 / 0.604 & 0.400 & 0.853 
& 0.393 & 0.859 & +180.7 & +7.8 \\
\rowcolor{gray!10}
\textbf{Category-Structured}
& \textbf{0.659 / 0.598} & \textbf{0.833} & \textbf{0.941} 
& \textbf{0.613 / 0.592} & \textbf{0.800} & \textbf{0.890} 
& \textbf{0.817} & \textbf{0.915} & \textbf{+483.9} & \textbf{+14.8} \\
\midrule
\multicolumn{11}{l}{\textit{Individual Adaptation (Built on Category-Structured Model)}} \\
\midrule
+ In-Context Learning 
& 0.616 / 0.486 & 0.848 & 0.634 
& 0.690 / 0.555 & 0.825 & 0.814 
& 0.837 & 0.724 & +497.9 & -9.2 \\
+ RLHF (DPO)
& 0.713 / 0.534 & \textbf{0.900} & \textbf{0.969} 
& 0.721 / 0.549 & \textbf{0.892} & 0.906 
& \textbf{0.896} & \textbf{0.938} & \textbf{+540.0} & \textbf{+17.7} \\
\rowcolor{gray!15}
\textbf{+ Steering} 
& \textbf{0.716 / 0.615} & \textbf{0.900} & 0.952 
& \textbf{0.745 / 0.608} & 0.850 & \textbf{0.921} 
& 0.875 & 0.936 & +525.0 & +17.4 \\
\bottomrule
\end{tabular}
}
\end{table}

\begin{figure*}[t]
\centering
\includegraphics[width=\linewidth]{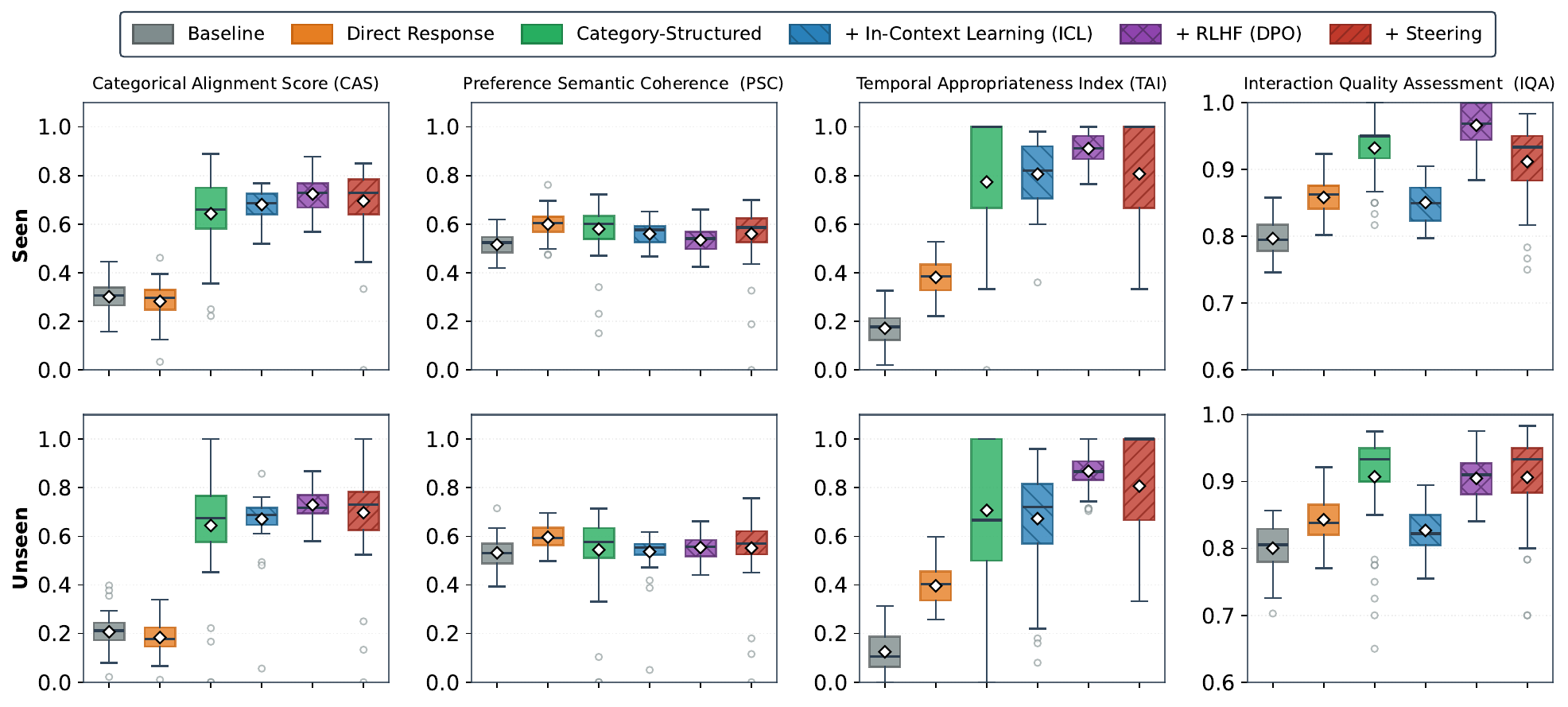}
\caption{Boxplots of Preference Understanding (CAS/PSC), Temporal Appropriateness Index (TAI), and Interaction Quality Assessment (IQA) across all methods for seen and unseen activity contexts.}
\label{fig:adaptation}
\end{figure*}

\subsection{Overall Performance Analysis (RQ3)}
\label{sec:overall_performance}

\textcolor{myblue}{We next evaluate whether the proposed two-stage framework improves proactive assistant behavior in the simulation setting. We separate this analysis into two parts: population-level learning and individual adaptation. The population-level analysis examines whether category-structured training provides a stronger cold start than an untuned model or direct-response fine-tuning. The individual-adaptation analysis examines whether user-specific feedback further improves behavior after the population model has been learned. Table~\ref{tab:comprehensive_comparison} provides a compact summary of average method performance, while Figure~\ref{fig:adaptation} visualizes the full context-level distribution of CAS, PSC, TAI, and IQA for seen and unseen activity contexts. For statistical reporting, we report means, standard deviations, and 95\% confidence intervals at the persona level. For method comparisons, we report corrected significance tests and effect sizes, with Bonferroni correction applied across planned comparisons.}

\noindent\textbf{Population-Level Learning.}
\textcolor{myblue}{We first examine whether category-structured population-level learning improves behavior beyond an untuned model. The Category-Structured model substantially improves timing alignment and interaction quality over the untuned baseline.}
Averaged across seen and unseen scenarios, TAI rises from 0.140 to 0.817 and IQA from 0.797 to 0.915 (Table~\ref{tab:comprehensive_comparison}). \textcolor{myblue}{Across the full simulated interaction distribution, the same pattern remains: Category-Structured improves TAI from 0.149 to 0.740 and IQA from 0.799 to 0.920 relative to the untuned baseline, with both improvements significant after correction ($p_{\mathrm{adj}} < 0.001$).} 
\textcolor{myblue}{The Direct Response model also improves over the baseline in the full interaction distribution (TAI 0.389, IQA 0.851), but its preference understanding remains substantially weaker (CAS 0.282 on seen contexts versus 0.643 for Category-Structured). The gains for Category-Structured are observed in both seen and held-out unseen activity contexts, suggesting that the model learns reusable preference structure rather than only memorizing training scenarios. However, since the evaluation is still simulation-based, we interpret these results as evidence of controlled generalization within our synthetic evaluation setting, not as proof of full real-world generalization.}

\textcolor{myblue}{Figure~\ref{fig:adaptation} further shows that these gains are visible across the full interaction distribution. The Category-Structured model has higher median TAI and IQA than both baselines, with lower variance in timing alignment. This pattern suggests that population-level structure provides a stronger cold start for proactive behavior, even before user-specific adaptation occurs. The gains are observed in both seen and held-out unseen activity contexts, suggesting controlled generalization within the synthetic evaluation setting. We do not interpret this as proof of full real-world generalization; rather, it shows that the learned preference structure transfers beyond the exact scenarios used in Stage~1 training.}

\textcolor{myblue}{PSC is reported in Table~\ref{tab:comprehensive_comparison} as a semantic-alignment diagnostic. Because PSC measures representation-level similarity between generated and reference preference descriptions, we interpret it together with CAS rather than as a standalone indicator of user benefit. The main behavioral evidence for population-level improvement comes from the combined gains in TAI and IQA, supported by improved CAS.}

\noindent\textbf{Individual Adaptation.}
\textcolor{myblue}{We next examine whether user-specific adaptation further improves behavior on top of the Category-Structured model. We compare In-Context Learning, RLHF with DPO, and our Activation Vector Steering method to test whether lightweight on-device steering can approach parameter-updating adaptation while keeping the model weights frozen.}
\textcolor{myblue}{In-Context Learning does not consistently improve interaction quality. Although it slightly reduces CAS variance, its IQA is significantly lower than the Category-Structured model in seen contexts ($\Delta=-0.082$, 95\% CI $[-0.097, -0.067]$, $p_{\mathrm{adj}}<0.001$, $d=2.19$). This suggests that retrieving recent interactions at inference time may introduce noise or overemphasize recent cases without providing stable personalization.}
\textcolor{myblue}{RLHF with DPO achieves the strongest overall mean performance in seen contexts (TAI $0.911 \pm 0.064$, IQA $0.966 \pm 0.030$) and serves as an upper-reference point when parameter updates are allowed. Activation Vector Steering remains below RLHF in seen contexts (TAI $0.807 \pm 0.234$; $\Delta_{\mathrm{TAI}}=-0.104$, $p_{\mathrm{adj}}=0.027$, $d=0.60$; IQA $0.912 \pm 0.056$; $\Delta_{\mathrm{IQA}}=-0.054$, $p_{\mathrm{adj}}<0.001$, $d=1.22$). However, this gap narrows substantially on unseen contexts, where Steering reaches TAI $0.807 \pm 0.234$ versus RLHF $0.868 \pm 0.076$ ($\Delta=-0.061$, $p_{\mathrm{adj}}=0.75$, not significant) and IQA $0.906 \pm 0.068$ versus RLHF $0.905 \pm 0.031$ ($\Delta=+0.001$, not meaningfully different). This pattern suggests that RLHF achieves stronger alignment on the specific activity distribution seen during training, while Steering generalizes more evenly to new contexts.}
\textcolor{myblue}{Compared with the non-adaptive Category-Structured model, Steering shows numerical improvements in TAI on unseen contexts ($\Delta=+0.100$, 95\% CI $[-0.004, +0.204]$) and equivalent IQA ($\Delta=-0.001$), though neither difference reaches the corrected significance threshold. These results indicate that Steering does not degrade the population-level model while providing on-device adaptation without parameter updates. Because Steering stores only small per-user activation directions and applies them at inference time, it provides a practical mechanism for individual alignment under mobile hardware and privacy constraints, approaching RLHF performance on held-out contexts at substantially lower deployment cost.}

\textcolor{myblue}{Figure~\ref{fig:adaptation} shows that Steering maintains stable IQA and TAI over the full interaction distribution, consistent with the long-horizon stability analysis in Section~\ref{sec:rq4}. This pattern indicates that the stored per-user steering directions retain feedback over time and continue to influence later responses without retraining.}

\begin{figure}[t]
    \centering
    \begin{minipage}[t]{0.45\columnwidth}
        \centering
        \includegraphics[width=\linewidth]{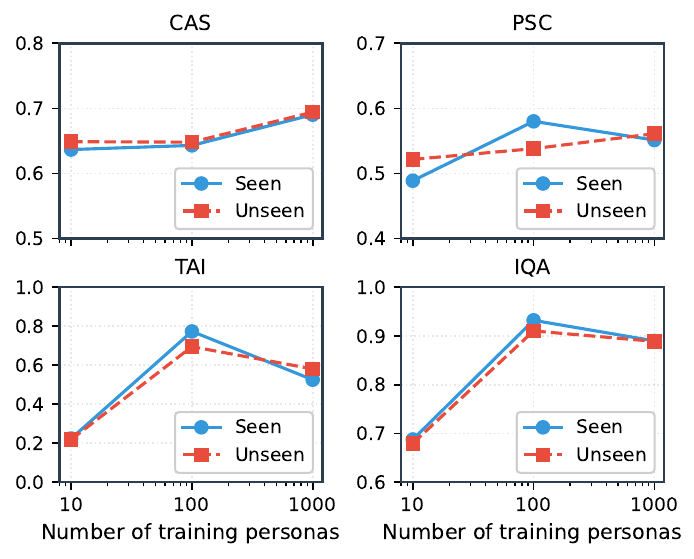}
        \vspace{-20pt}
        \caption{Scaling of population-level learning with training diversity.
        Performance as the number of training personas increases (10, 100, 1000), shown for seen and unseen activity contexts.}
        \vspace{-10pt}
        \label{fig:scaling}
    \end{minipage}
    \hspace{0.04\columnwidth}
    \begin{minipage}[t]{0.45\columnwidth}
        \centering
        \includegraphics[width=\linewidth]{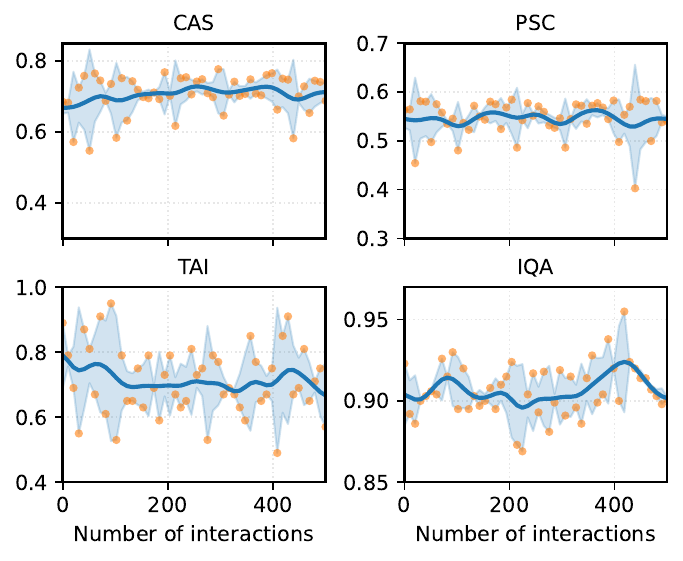}
        \vspace{-20pt}
        \caption{Long-horizon adaptation over 500 interactions. All four metrics stay 
        stable with natural fluctuations, showing no clear degradation over the simulated horizon.}
        \label{fig:activity_scaling}
    \end{minipage}
\end{figure}

\subsection{Population Diversity and Adaptation Dynamics (RQ4)}
\label{sec:rq4}
We next study how population-level training diversity and long-term interaction depth influence preference understanding, timing, and overall interaction quality.

\noindent\textbf{Scaling with training diversity.}
We vary the number of personas used in Stage~1 training while fixing the number of activities per persona to 100 ($10{\times}100$, $100{\times}100$, and $1000{\times}100$). Figure~\ref{fig:scaling} reports CAS, PSC, TAI, and IQA for both seen and unseen activity contexts as the number of training personas increases. \textcolor{myblue}{The largest improvement occurs from 10 to 100 personas, especially for TAI and IQA, suggesting that timing decisions and interaction quality benefit from broader preference coverage. CAS remains relatively stable, while PSC changes only modestly and does not follow a strictly monotonic trend.}

\textcolor{myblue}{Increasing from 100 to 1,000 personas does not produce further gains across all metrics; TAI and IQA slightly decrease, while CAS and PSC remain mostly stable. One possible explanation, consistent with our dataset sanity check, is that increasing the generated dataset size does not necessarily add proportional activity or preference diversity. After moderate population coverage is reached, additional personas may contribute repeated or highly similar preference expressions rather than substantially new training signal. We therefore interpret the 100-to-1,000 trend as a saturation pattern, not as evidence that larger synthetic populations are always better. The similar trends for seen and unseen contexts suggest controlled generalization within our generated data distribution, rather than unrestricted real-world generalization.}

\noindent\textbf{Scaling with interaction depth.}
To assess stability over long-term use, we extend the interaction horizon to 500 proactive interactions per simulated persona while applying Activation Vector Steering. Figure~\ref{fig:activity_scaling} plots CAS, PSC, TAI, and IQA over time. \textcolor{myblue}{CAS, PSC, and IQA remain relatively stable with natural fluctuations as activity contexts change, while TAI shows larger variance because timing decisions are more sensitive to moment-level interruptibility.}
We do not observe clear drift towards overly conservative or overly aggressive behavior, \textcolor{myblue}{nor a sustained decline in interaction quality under the simulated setting. This suggests that the steering mechanism can retain useful feedback over extended use without causing obvious instability, although the result should be interpreted as a simulation-based stability check rather than proof of long-term real-world robustness.} The pattern is consistent with our user study, where participants report \textcolor{myblue}{early adaptation followed by more predictable assistant behavior.}

\begin{table*}[t]
\small
\setlength{\tabcolsep}{4pt}
\renewcommand{\arraystretch}{1.08}

\begin{minipage}[t]{0.485\textwidth}
\vspace{0pt}

\captionsetup{
    font=small,
    justification=raggedright,
    singlelinecheck=false,
    skip=4pt
}

\captionof{table}{
Performance across assistant model architectures using GPT-4.1 as the user simulator.
}
\label{tab:assistant_ablation}

\begin{tabular*}{\linewidth}{
    @{\extracolsep{\fill}}
    lcccc
    @{}
}
\toprule
\textbf{Assistant Model}
& \textbf{CAS}
& \textbf{PSC}
& \textbf{TAI}
& \textbf{IQA} \\
\midrule
\rowcolor{gray!20}
Llama-3.2-3B & 0.731 & 0.595 & 0.924 & 0.922 \\
Llama-3.2-1B & 0.742 & 0.557 & 0.545 & 0.917 \\
Qwen3-1.7B   & 0.024 & 0.021 & 0.113 & 0.838 \\
Qwen3-0.6B   & 0.546 & 0.437 & 0.243 & 0.808 \\
\bottomrule
\end{tabular*}

\end{minipage}
\hfill
\begin{minipage}[t]{0.485\textwidth}
\vspace{0pt}

\captionsetup{
    font=small,
    justification=raggedright,
    singlelinecheck=false,
    skip=4pt
}

\captionof{table}{
Generalization of the trained Llama-3.2-3B assistant across user simulators. Metrics are averaged over seen and unseen activity contexts.
}
\label{tab:user_generalization}

\begin{tabular*}{\linewidth}{
    @{\extracolsep{\fill}}
    lcccc
    @{}
}
\toprule
\textbf{User Model}
& \textbf{CAS}
& \textbf{PSC}
& \textbf{TAI}
& \textbf{IQA} \\
\midrule
\rowcolor{gray!20}
GPT-4.1         & 0.730 & 0.595 & 0.875 & 0.923 \\
Llama-3.1-70B   & 0.449 & 0.350 & 0.404 & 0.946 \\
Claude-Sonnet-4 & 0.707 & 0.550 & 0.546 & 0.619 \\
\bottomrule
\end{tabular*}

\end{minipage}
\end{table*}

\subsection{Cross-Model Validation (RQ5)}
\label{sec:cross_model_validation}
Finally, we examine whether the observed trends depend on a specific assistant backbone or a specific user simulator. This analysis is intended as a robustness check rather than a full benchmark across model families.

\noindent\textbf{Assistant Model Comparison.}
Table~\ref{tab:assistant_ablation} reports CAS, PSC, TAI, and IQA when the two-stage pipeline is applied to several compact assistant backbones while keeping the user simulator fixed to GPT-4.1. Llama-3.2-3B achieves the most balanced performance, with strong preference understanding, timing alignment, and interaction quality (CAS 0.731, PSC 0.595, TAI 0.924, IQA 0.922). 
This supports our use of a 3B Llama-based model as a practical trade-off between proactive interaction quality and on-device feasibility. Llama-3.2-1B reaches comparable CAS and IQA, but much lower TAI (0.545), suggesting that smaller models may recognize preference categories while struggling with fine-grained timing decisions. 

\textcolor{myblue}{The Qwen models perform less consistently under the same training and prompting setup. In particular, Qwen3-1.7B performs worse than Qwen3-0.6B on CAS and PSC. Manual inspection suggests that this is largely due to missing or empty structured reasoning fields: many Qwen3-1.7B outputs have missing or empty fields for detailed preference reasons, whereas Qwen3-0.6B produces such empty fields less often. Because CAS and PSC depend on whether the model produces category-aligned structure and semantically meaningful preference descriptions, these missing fields directly lower the scores. We therefore interpret this result as evidence that model family and instruction-following behavior affect structured preference learning, rather than as a simple parameter-scaling effect.}

\noindent\textbf{User Simulator Generalization.}
Table~\ref{tab:user_generalization} examines how the trained Llama-3.2-3B assistant behaves when paired with different user simulators. With GPT-4.1 as the user simulator, the assistant obtains the highest CAS, PSC, and TAI, which is expected because GPT-4.1 is also used in the main data construction pipeline.
When switching to Llama-3.1-70B, \textcolor{myblue}{CAS, PSC, and TAI decrease, while IQA increases. This mismatch suggests that the simulator's rating behavior and preference-expression style differ from GPT-4.1: the assistant may receive high perceived-quality ratings even when its structured preference predictions align less closely with the reference labels.}
With Claude-Sonnet-4, CAS and PSC remain relatively high, but IQA drops substantially, \textcolor{myblue}{indicating a different evaluation style or stricter judgment of interaction quality.}
Across these simulator changes, \textcolor{myblue}{the framework remains applicable, but the absolute metric values vary substantially across user models. }
We therefore treat simulator-based results as controlled comparisons \textcolor{myblue}{for stress-testing method behavior, not as absolute estimates of real-user quality. This further motivates the human-subject study, which complements simulator-based evaluation with direct user feedback.}

\section{User Study}
\label{sec:user_study}

We conducted two complementary user studies to evaluate how well the system supports preference-aligned interactions: (1) a preference detection study testing whether participants can distinguish personalized from non-personalized responses, and (2) an interactive adaptation study examining how real-time personalization shapes satisfaction and decision patterns across multi-session interactions. The study was approved by our institutional review board, and all participants provided informed consent prior to participation. \textcolor{myblue}{Detailed user study materials, condition definitions, questionnaire items, and interview prompts are provided in Appendix~\ref{sec:appendix_interface}.}

\subsection{Preference Detection Study}
\noindent\textbf{Participants.}
We recruited \textcolor{myblue}{34} participants (14 female, \textcolor{myblue}{20} male, age range 18--54) through university mailing lists and online community boards. Participants had varying levels of experience with smart assistants, ranging from no prior experience (n=12) to daily use (n=2), as shown in Table~\ref{tab:demographics}. Each participant received \$15 compensation for approximately one hour of participation.

\noindent\textbf{Study Design.}
We created 40 storyboard pairs that describe common daily situations 
\textcolor{myblue}{including morning routines, work activities, leisure time, and evening contexts. Each pair contained two assistant responses to the same situation: one generated by our personalized two-stage assistant and one generated by a non-personalized baseline. The two responses differed in one or more preference dimensions, most often timing, autonomy, and communication style. Each participant evaluated 10 randomly sampled pairs. For each pair, participants selected the response that ``feels more appropriate'' for the situation and briefly explained their choice. The order of personalized and non-personalized responses was randomized to reduce position bias.}

\noindent\textbf{Results.}
Participants preferred the personalized response in 62\% of all comparisons. \textcolor{myblue}{This descriptive result suggests that personalization was detectable even in short, text-only scenarios where participants did not know the underlying synthetic personas or their preference histories.}
\textcolor{myblue}{Participants' explanations show that they often preferred a response because it better matched the timing, context, autonomy expectation, or communication style of the situation.}
They rejected suggestions that came ``too early,'' ``too late,'' or during an inappropriate activity, even if the content of the suggestion was reasonable. 
\textcolor{myblue}{They also preferred responses that preserved control, such as asking before taking action, and responses that matched the desired level of detail.}
\textcolor{myblue}{These comments support the preference structure used in our framework. Participants often cared about the same broad dimensions in Table~\ref{tab:preference_categories} but the preferred direction differed across individuals. This suggests that population-level learning can identify common preference dimensions, but cannot fully determine how each user wants those dimensions to be applied. Individual adaptation is therefore needed to learn the user's preferred direction within each dimension.}

\subsection{Interactive Adaptation Study}
\noindent\textbf{Study Design.}
\textcolor{myblue}{The same 34 participants completed an interactive adaptation study immediately after the preference detection study.} This part of the study examines how real-time personalization influences users’ perceptions during a series of interactions. 
\textcolor{myblue}{Participants completed 10 storyboard-based proactive assistant interactions covering morning, noon, evening, work-related, and leisure scenarios.}
After each assistant action, participants rated five aspects of the interaction: timing appropriateness, intrusiveness, content value, contextual relevance, and autonomy respect. 
\textcolor{myblue}{They could also provide short text feedback through the study interface, with researchers assisting only in entering the feedback.}

\textcolor{myblue}{We used a Static--Adaptive--Mixed (SAM) condition design.} Participants were randomly assigned to one of three approximately balanced between-subject conditions: 
\textcolor{myblue}{(S) a \emph{Static} baseline assistant with fixed population-average preference weights, (A) an \emph{Adaptive} assistant that updated activation-based steering after each feedback round, and (M) a \emph{Mixed} assistant that alternated between adaptive responses on odd-numbered interactions and static responses on even-numbered interactions.} 
\textcolor{myblue}{The Mixed condition was included for two reasons. First, it reduces demand effects by avoiding a simple adaptive-versus-static comparison, which could allow participants to infer the study hypothesis and bias their responses. By interleaving adaptive and non-adaptive behaviors within the same sequence, the study presents a less predictable interaction pattern. Second, it enables us to examine whether participants perceive adaptation as a sequence-level behavior, rather than as isolated improvements at individual interaction points.}
To reduce expectation bias, \textcolor{myblue}{we described the study as comparing different "pre-configured assistant personalities" rather than explicitly labeling conditions as adaptive or non-adaptive.}

After completing all 10 interactions, \textcolor{myblue}{participants completed a post-study questionnaire measuring perceived overall change, perceived adaptation, trust, comfort and willingness to reuse, and overall satisfaction. They also took part in a short semi-structured interview about how the assistant's behavior changed across the interaction sequence.}

\begin{figure}[t]
\centering
\begin{minipage}[h]{0.43\columnwidth}
\raggedright
\captionof{table}{Participant demographics \textcolor{myblue}{($N=34$)}.}
\label{tab:demographics}
\small
\setlength{\tabcolsep}{6pt}
\begin{tabular}{llc}
\toprule
\textbf{Category} & \textbf{Group} & \textbf{Count} \\
\midrule
\multirow{3}{*}{Age range}& 18--24 & 14\\
 & 25--34 & 15\\
 & 45--54 & 5 \\
\midrule
\multirow{2}{*}{Gender}
 & Female & 14 \\
 & Male & 20\\
\midrule
\multirow{6}{*}{Education}
 & Some college (no degree) & 9 \\
 & High school diploma & 1 \\
 & Bachelor’s degree & 4 \\
 & Master’s degree & 14\\
 & Doctoral degree & 6\\
\midrule
\multirow{4}{*}{\makecell[l]{Smart assistant\\experience}} & None & 12\\
 & Occasional user & 14\\
 & Regular user & 6 \\
 & Daily user & 2 \\
\bottomrule
\end{tabular}
\end{minipage}
\hfill
\begin{minipage}[h]{0.48\columnwidth}
\centering
\includegraphics[width=\linewidth]{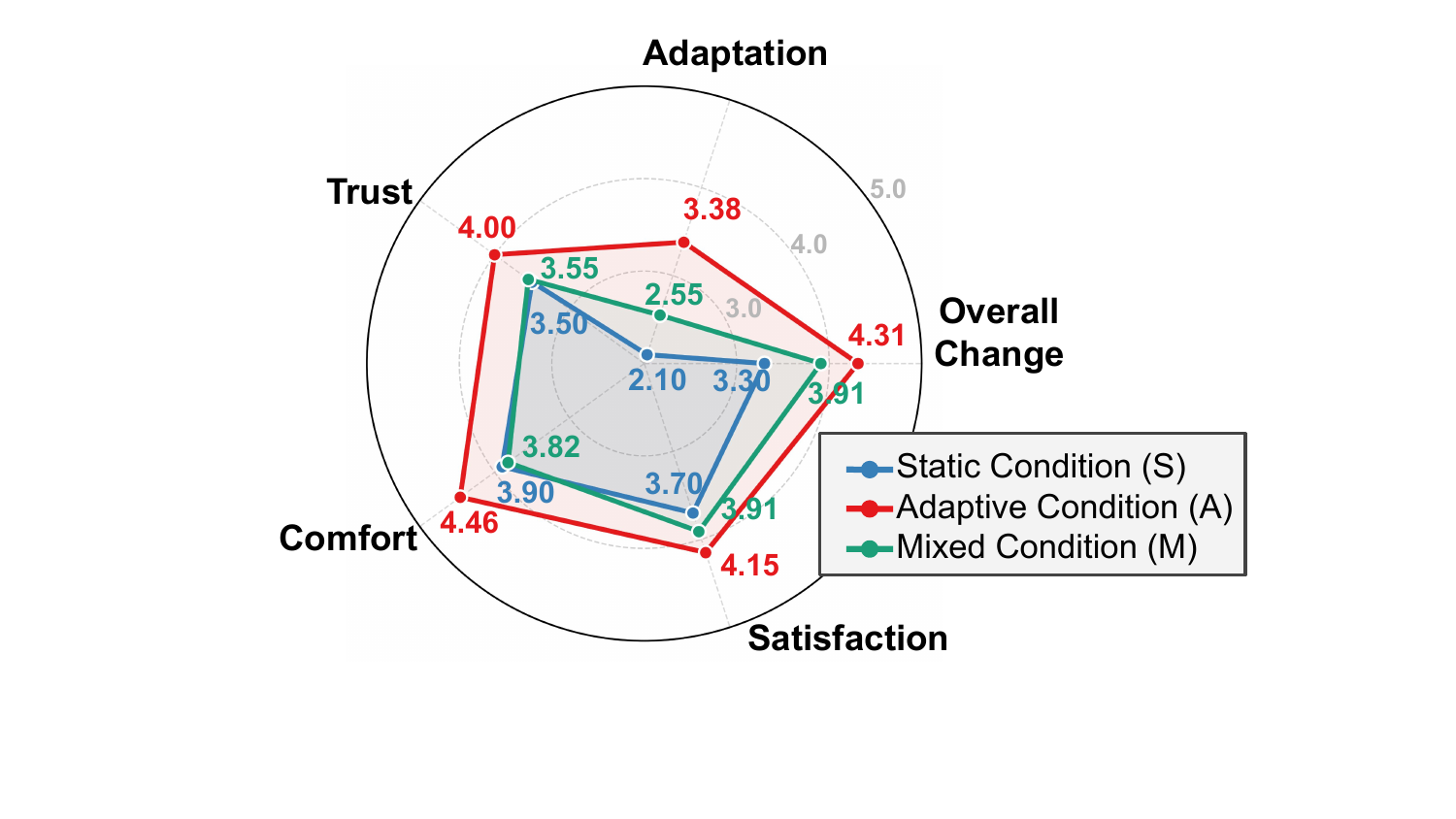}
\captionof{figure}{User perception of proactive adaptation across conditions.}
\label{fig:radar}
\end{minipage}
\end{figure}

\noindent\textbf{Results.}
As summarized in Figure~\ref{fig:radar}, the adaptive condition received higher mean ratings than the static baseline on all five subjective items: perceived overall change, perceived adaptation, trust, comfort and willingness to reuse, and overall satisfaction. 
\textcolor{myblue}{Participants in the adaptive condition often described the assistant as using earlier feedback, adjusting quickly, or becoming less disruptive over time. In contrast, participants in the Static condition described the assistant as consistent but unchanging, and several noted that their ratings did not seem to affect later behavior.}

\textcolor{myblue}{The Mixed condition provides additional evidence that participants noticed adaptation as a sequence-level behavior. Its ratings generally fell between the Static and Adaptive conditions, and several participants could identify which sessions reflected their earlier feedback and which did not. This suggests that adaptation is not only reflected in individual response ratings, but is also visible to users as a pattern across interactions. For proactive assistants, this matters because inconsistent adaptation may weaken trust even when individual responses are reasonable. Participants' comments further show which forms of adaptation mattered most. Common examples included fewer proactive interruptions during focused work, more concise messages after participants requested brief updates, and more cautious behavior when participants preferred greater control. These examples align with the preference dimensions used in our framework, especially timing, autonomy, communication style, and contextual relevance. They also match the early adaptation trends observed in the simulation results, where steering improved during the first several interactions.}

\textcolor{myblue}{Overall, the user study suggests that people value proactive assistants that learn in a stable and correctable way. Participants judged not only whether a single response was useful, but also whether the assistant appeared to incorporate feedback and become more predictable over time. This finding supports the role of individual adaptation in our framework and shows why response quality alone is not sufficient for proactive assistant evaluation.}

\noindent\textbf{Power Analysis and Interpretation.}
\textcolor{myblue}{We conducted a power analysis to contextualize the human-study sample size. The Preference Detection Study uses a binary A/B choice design, where the main question is whether participants select the personalized response above chance. Because each participant completed multiple A/B comparisons, we treat the participant, rather than each individual choice, as the independent unit for this power interpretation. For a one-sample proportion test against chance level ($p_0=0.5$), a medium effect size (Cohen's $h=0.5$, approximately corresponding to $p_1 \approx 0.74$), $\alpha=.05$, and power $=.80$ require approximately 32 independent participant-level observations. Our updated sample of 34 participants therefore reaches the conventional sample-size requirement for detecting a medium A/B preference effect. Because the observed aggregate preference rate was 62\%, we interpret the detection-study result as descriptive evidence that personalization was noticeable in these short scenarios, rather than as a definitive estimate of a population-level effect.}

\textcolor{myblue}{At the same time, the Interactive Adaptation Study compares Static, Adaptive, and Mixed conditions in a between-subject design. This part would require a substantially larger sample to detect medium between-condition effects with conventional power. We therefore interpret the adaptation-study results descriptively, using them as complementary user-facing evidence of perceived adaptation, trust, comfort, and satisfaction rather than as definitive population-level hypothesis tests. Larger studies with sufficient statistical power and longer in-the-wild deployments remain necessary to estimate robust real-world effect sizes.}

\section{Findings, Discussion, and Implications}
\label{sec:discussion}

\subsection{Key Findings}

\textcolor{myblue}{Our results show that large-scale simulated persona interactions can provide a useful population-level starting point for proactive assistant learning, but that this benefit depends on the behavioral diversity of the generated traces. In Stage~1, category-structured training improves timing alignment and preference understanding compared with untuned and response-only baselines, suggesting that synthetic interactions can help models learn reusable patterns in how users express preferences. However, the scaling results show that increasing the number of generated personas does not automatically produce proportional gains, indicating that a larger synthetic cohort is most useful when it introduces new variation in activities, contexts, preference expressions, and feedback behavior rather than repeating similar preference templates across superficially different personas. The results also support our population-to-individual design: users often refer to similar preference dimensions, including timing, autonomy, communication style, domain priority, and context adaptation, but they may prefer different behaviors within the same dimension. Population-level learning can therefore provide a strong initial model, while individual adaptation remains necessary for aligning proactive behavior with each user's expectations. The human study further suggests that users evaluate the assistant as an adaptive system rather than as isolated responses, since participants noticed behavioral change quickly and responded positively when the assistant appeared to incorporate prior feedback, even when the adapted behavior was not always ideal.}

\subsection{Discussion}

\textcolor{myblue}{These findings clarify both the value and the limits of simulation for proactive assistant learning. Simulation can provide structured, repeatable, and privacy-preserving interaction data before real deployment, making it useful as an early-stage bridge for training and evaluation. At the same time, our results do not support treating simulation as a substitute for real-user evaluation. The saturation observed in the scaling analysis suggests that adding more synthetic personas may not improve learning if the generated traces do not add new forms of preference variation. This limitation is especially important for proactive daily-life assistants, because user expectations in this setting are still developing: many users do not yet have a stable mental model of how a proactive assistant should behave, how much initiative it should take, or how they should correct it when it acts poorly. As a result, even real-user data collected today may reflect early and exploratory behavior rather than stable long-term preferences. This perspective also helps interpret the 62\% preference rate in the detection study. Rather than treating it as a direct measure of model accuracy, we view it as evidence that some proactive preferences are observable in short, controlled scenarios, while other preferences remain ambiguous, context-dependent, or difficult to express through stated responses alone.}

\textcolor{myblue}{The human sanity check provides additional support for the plausibility of the generated training data, but its scope should also be interpreted carefully. Across 1,000 ratings from five raters, 92.3\% of persona--activity--preference examples were judged usable or borderline, with strong scores for context relevance, internal consistency, and realism. This suggests that the generated examples were generally coherent with the provided persona profiles and activity contexts. However, raters inspected static examples rather than interacting with an adaptive assistant over time, so this check screens for plausibility rather than long-term behavioral fidelity. LLM-generated personas may also reflect systematic biases, including stereotyped links between demographic attributes and behavior, overly polished language, or repeated preference templates. These limits are mirrored in the adaptation setting: feedback is not only an optimization signal, but also part of how users judge whether the assistant is responsive. Ignored suggestions are especially difficult to interpret because they provide a signal without an explanation; silence may reflect rejection, distraction, low urgency, social context, or a preference that the user has not yet expressed in words. Our current steering method treats ignored suggestions as negative evidence, which is a practical approximation, but future systems need richer contextual signals to distinguish among these cases. The steering mechanism itself also raises open questions about stability and preference conflict. Because inference-time steering combines multiple category-specific directions through linear addition, conflicting preference dimensions may partially offset one another in activation space. Although our current design reduces this risk through bounded steering strengths, decay for unsupported signals, and recent-feedback-based updates, future work should test single-dimension, paired-dimension, and all-dimension steering conditions to better understand category interactions and long-term stability.}

\subsection{Implications and Future Work}

\textcolor{myblue}{For proactive assistant design, these results suggest that personalization should be correctable and legible, not only accurate. Users may be uncomfortable with an assistant that appears to infer too much too soon or adapts in ways that are fast but opaque. A more appropriate design goal is to let the assistant improve gradually while giving users lightweight ways to correct its behavior, especially in daily-life settings where preferences may be incomplete, unstable, or revealed only through repeated interaction. Future simulation and deployment pipelines should therefore model noisy, sparse, delayed, and conflicting feedback, since real users may ignore assistants, provide partial comments, change their preferences over time, or express different needs across timing, autonomy, and communication style. Beyond language, multimodal context may help systems interpret the gap between what users say and what they do; for example, understanding whether an ignored suggestion occurred while the user was physically occupied, socially engaged, moving through a space, or simply uninterested could improve feedback interpretation. Vision and other sensor-based modalities may help capture such behavioral context, but they also introduce privacy and consent challenges, so future work should study multimodal preference learning together with privacy-preserving sensing and local processing.}

\textcolor{myblue}{The broader implication is that our framework transfers most directly at the level of preference modeling and adaptation, not as fixed rules for one modality. Proactive systems across mobile notifications, wearable alerts, in-car assistants, smart-home displays, and embodied agents all face similar structural questions: when to act, how visible the action should be, how much control the user keeps, and how the system adapts after feedback. However, each modality changes the meaning of interruption and control. A silent phone notification may be acceptable when a spoken interruption would be intrusive; a wearable cue may preserve privacy but provide limited feedback bandwidth; an embodied agent may introduce spatial and social expectations that are absent in text or voice. Therefore, timing thresholds, intrusiveness judgments, and feedback channels should be recalibrated for each interaction form. Finally, future evaluation should move beyond storyboard-based judgments toward participant-specific activities and longer in-the-wild deployments, since our current human study captures perceived appropriateness and adaptation under controlled conditions but does not fully capture attention, emotion, social setting, activity pressure, or delayed reactions during daily use.}

\section{Conclusion}
By generating large-scale interaction data with diverse personas and daily contexts, we address the challenge of collecting long-term preference signals in real deployments. Our two-stage approach first learns population-level preference patterns from synthetic data and then adapts to each user through lightweight activation steering during inference. Experiments with 1{,}000 simulated personas and a complementary user study indicate improvements in preference understanding, timing decisions, and user-perceived interaction quality. These findings suggest that combining synthetic population-level learning with efficient on-device adaptation offers a practical path toward personalized, privacy-preserving proactive assistants.

\bibliographystyle{ACM-Reference-Format}
\bibliography{references}

\appendix
\section{End-to-End Training Example of Synthetic Preference Supervision}
\label{appendix:example}


\begin{tcolorbox}[title=An Example of Synthetic Preference]

\textbf{(1) Persona Profile.} \\
\textit{Persona ID:} P042 (UUID: 2f303f19-17d6-4db9-9c7b-fda0154ab5c7) \\
\textit{Demographics:} 52-year-old male; married; associate’s degree; human
resources worker; resides in Queen Creek, AZ, USA. \\
\textit{Lifestyle and background:} Family-oriented and highly structured yet
flexible; actively coaches his son’s soccer team; organizes family game nights;
bilingual (English/Spanish); experienced in mediation and conflict resolution;
values organization, efficiency, and predictable routines.
This persona profile is sampled from a census-aligned synthetic persona pool and
conditions daily routines and interaction behavior. It does not encode explicit
preference categories.

\vspace{0.6em}
\textbf{(2) Activity Context.} \\
\textit{Timeframe:} 6:30--6:55 AM \\
\textit{Activity:} Morning hygiene routine \\
\textit{Category:} Health \& Wellness \\
\textit{Context summary:} The persona has just woken up and is progressing
through a private, structured morning routine involving showering and grooming,
with limited tolerance for interruptions.
This activity context represents a potential proactive intervention window and
contains no preference annotations.

\vspace{0.6em}
\textbf{(3) Active Preference Categories.} \\
Based on the simulated persona’s reflection in this context, the following
subset of preference categories is identified as active:
\begin{itemize}
    \item Scheduling Preference (Timing and Interruption)
    \item Communication Style
    \item Context Adaptation
\end{itemize}
Only this subset is relevant in the current context; other categories are not
expressed or elaborated.

\vspace{0.6em}
\textbf{(4) Natural-Language Preference Description.} \\
The simulated persona expresses the following preferences in natural language:
\begin{quote}
In the mornings, I prefer to move through my routine without distractions or
unnecessary notifications. If a reminder is urgent, it should wait until I am
out of the shower and nearly finished. Communication should be calm, friendly,
and concise, helping me start the day smoothly without unnecessary chatter. The
assistant should adapt to the structured and efficiency-focused nature of my
morning routine.
\end{quote}

\vspace{0.6em}
\textbf{(5) Preferred Assistant Response.} \\
The simulated persona provides the following assistant response as appropriate
and aligned with the stated preferences:
\begin{quote}
Good morning, Wes. You’re moving smoothly through your routine. Just a quick
heads-up: there’s a slight change in your calendar regarding your son’s soccer
practice—it’s been moved to 5:00 PM today. Have a great day, and feel free to
check in once you’re all set.
\end{quote}

\end{tcolorbox}

\newpage

\section{Persona Sampling Validity Details}
\label{app:persona_sampling_validity_details}
\begin{center}
\footnotesize
\setlength{\tabcolsep}{6pt}
\renewcommand{\arraystretch}{1.05}
\begin{longtable}{p{0.36\linewidth}rrr}
\caption{Detailed marginal distribution comparison between the sampled 1,000 personas and the 100k adult source snapshot.}
\label{tab:persona_sampling_detailed}\\
\toprule
\textbf{Category} & \textbf{Source (\%)} & \textbf{Sample (\%)} & \textbf{$|\Delta|$ (pp)} \\
\midrule
\endfirsthead

\toprule
\textbf{Category} & \textbf{Source (\%)} & \textbf{Sample (\%)} & \textbf{$|\Delta|$ (pp)} \\
\midrule
\endhead

\midrule
\multicolumn{4}{r}{Continued on next page} \\
\midrule
\endfoot

\bottomrule
\endlastfoot

\multicolumn{4}{l}{\textbf{Age Group} (JS=0.032, TV=0.036)} \\
18--24 & 11.66 & 12.70 & 1.04 \\
25--34 & 17.46 & 18.40 & 0.93 \\
35--44 & 16.66 & 16.20 & 0.46 \\
45--54 & 15.72 & 17.30 & 1.58 \\
55--64 & 16.60 & 15.30 & 1.30 \\
65+ & 21.90 & 20.10 & 1.80 \\

\addlinespace[0.3em]
\multicolumn{4}{l}{\textbf{Gender} (JS=0.014, TV=0.017)} \\
Female & 51.13 & 52.80 & 1.67 \\
Male & 48.87 & 47.20 & 1.67 \\

\addlinespace[0.3em]
\multicolumn{4}{l}{\textbf{Education} (JS=0.048, TV=0.037)} \\
$<$9th grade & 4.46 & 6.20 & 1.74 \\
9th--12th, no diploma & 6.45 & 5.70 & 0.75 \\
High school & 27.25 & 27.40 & 0.15 \\
Some college & 21.18 & 22.60 & 1.42 \\
Associate's & 8.40 & 8.80 & 0.40 \\
Bachelor's & 20.16 & 19.50 & 0.66 \\
Graduate & 12.09 & 9.80 & 2.29 \\

\addlinespace[0.3em]
\multicolumn{4}{l}{\textbf{Marital Status} (JS=0.020, TV=0.014)} \\
Never married & 31.70 & 31.80 & 0.10 \\
Married & 48.78 & 49.70 & 0.92 \\
Divorced & 11.53 & 10.10 & 1.43 \\
Widowed & 6.14 & 6.40 & 0.26 \\
Separated & 1.84 & 2.00 & 0.15 \\

\addlinespace[0.3em]
\multicolumn{4}{l}{\textbf{Occupation Group} (JS=0.028, TV=0.025)} \\
Not in workforce & 30.41 & 31.20 & 0.79 \\
Mgmt./Business/STEM/Arts & 14.50 & 14.60 & 0.10 \\
Edu./Healthcare/Community & 7.16 & 6.00 & 1.16 \\
Sales/Office/Admin & 13.79 & 13.20 & 0.59 \\
Service & 5.32 & 4.90 & 0.42 \\
Construction/Maint./Natural Res. & 5.16 & 5.30 & 0.14 \\
Production/Transport/Material & 5.74 & 5.40 & 0.34 \\
Other & 17.91 & 19.40 & 1.49 \\

\addlinespace[0.3em]
\multicolumn{4}{l}{\textbf{Region} (JS=0.037, TV=0.034)} \\
Northeast & 17.24 & 15.70 & 1.54 \\
Midwest & 20.50 & 19.00 & 1.50 \\
South & 37.87 & 38.30 & 0.44 \\
West & 23.33 & 26.30 & 2.97 \\
Territory & 1.06 & 0.70 & 0.36 \\

\addlinespace[0.3em]
\multicolumn{4}{l}{\textbf{Top States + Other} (JS=0.046, TV=0.035)} \\
CA & 11.60 & 13.30 & 1.70 \\
TX & 8.48 & 9.10 & 0.62 \\
FL & 6.70 & 6.50 & 0.20 \\
NY & 5.86 & 4.60 & 1.26 \\
IL & 3.81 & 4.50 & 0.68 \\
PA & 3.98 & 3.50 & 0.48 \\
OH & 3.51 & 2.50 & 1.01 \\
GA & 3.16 & 2.80 & 0.36 \\
NC & 3.15 & 3.40 & 0.25 \\
MI & 3.01 & 2.80 & 0.21 \\
Other & 46.73 & 47.00 & 0.26 \\

\end{longtable}
\end{center}

\begin{figure}[h]
\centering
\vspace{-6pt}
\includegraphics[width=0.75\linewidth,trim=10 10 10 10,clip]{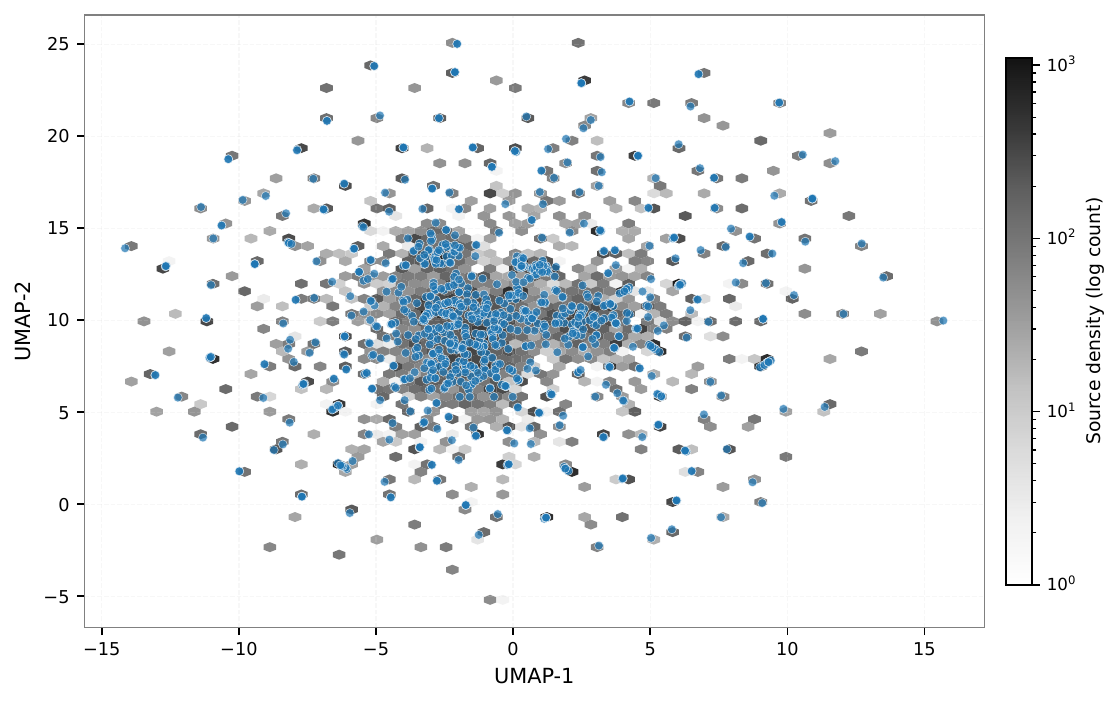}
\vspace{-6pt}
\caption{UMAP of persona-text embeddings}
\vspace{-6pt}
\label{fig:persona_umap}
\end{figure}

\section{User Study Interfaces and Materials}
\label{sec:appendix_interface}

This appendix provides additional examples of the user study interfaces and materials used across the Preference Detection Study and the Interactive Adaptation Study. These materials are included to clarify the structure of the storyboard-based tasks, the Static--Adaptive--Mixed (SAM) condition design, and the types of preference signals collected from participants. They are provided for methodological transparency and do not present additional experimental results.

Participants were presented with short storyboard sequences depicting everyday activities, accompanied by a textual description of the situation and one or two proactive assistant responses. Depending on the study phase, participants either compared alternative assistant behaviors or provided feedback on a single delivered interaction. The interface recorded participants' choices, Likert-scale ratings, optional free-form comments, and structured feedback related to the preference dimensions used in our framework, including timing, communication style, autonomy, context adaptation, and domain priorities.

Figure~\ref{fig:study_c} shows an example storyboard-based task used in the Preference Detection Study, where participants compared two assistant responses following the same activity context. Figure~\ref{fig:study_b} shows an example interface used in the Interactive Adaptation Study, where participants evaluated a delivered assistant response and provided preference feedback that could be used for adaptation depending on the assigned condition.

\begin{figure}[h]
\centering
\vspace{-6pt}
\includegraphics[width=0.75\linewidth,trim=10 10 10 10,clip]{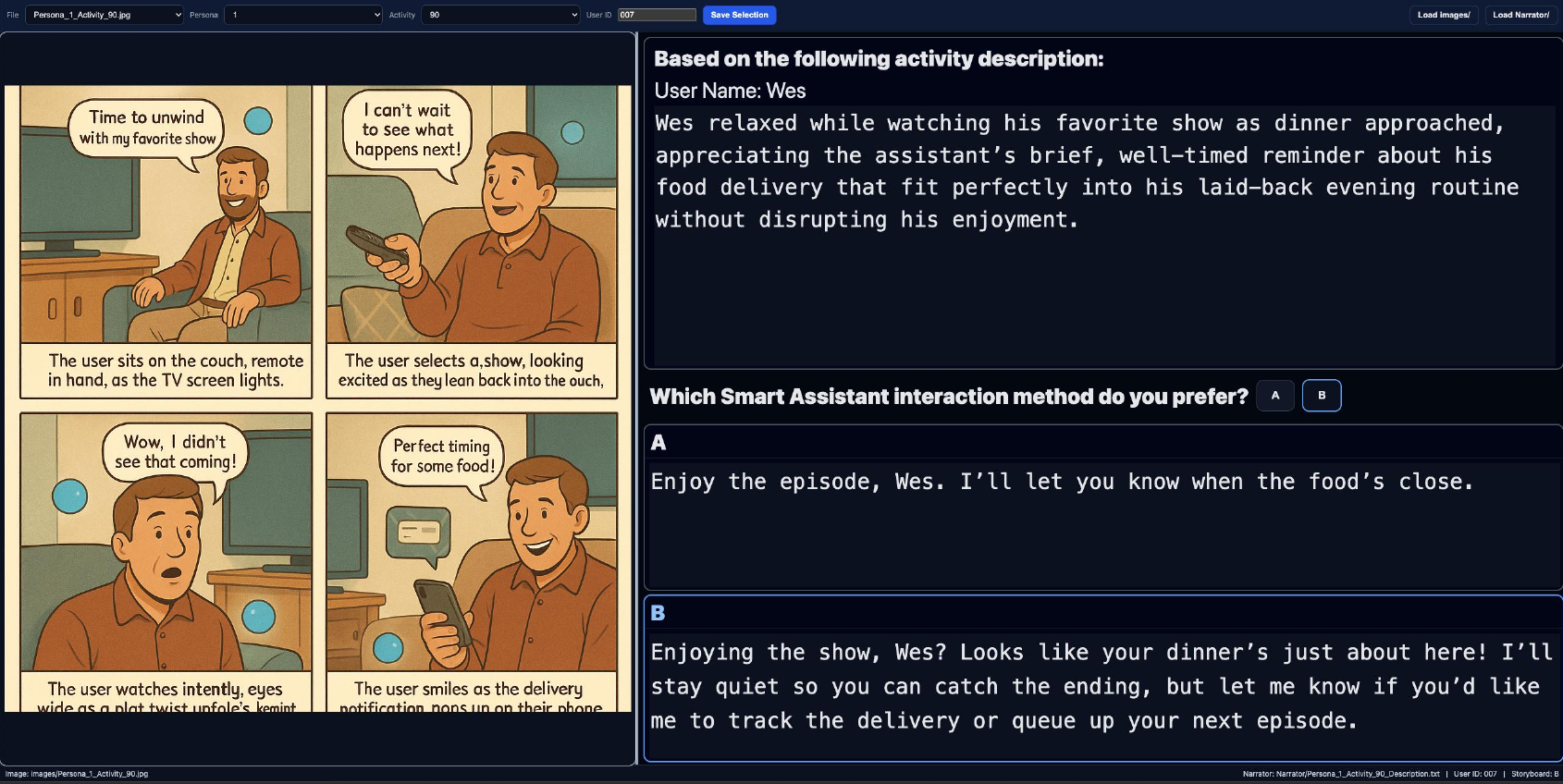}
\vspace{-6pt}
\caption{Example of a storyboard-based preference comparison task used in the Preference Detection Study. Participants selected the assistant response that felt more appropriate for the described situation and provided a short explanation.}
\vspace{-6pt}
\label{fig:study_c}
\end{figure}

\begin{figure}[h]
\centering
\vspace{-6pt}
\includegraphics[width=0.75\linewidth,trim=10 10 10 10,clip]{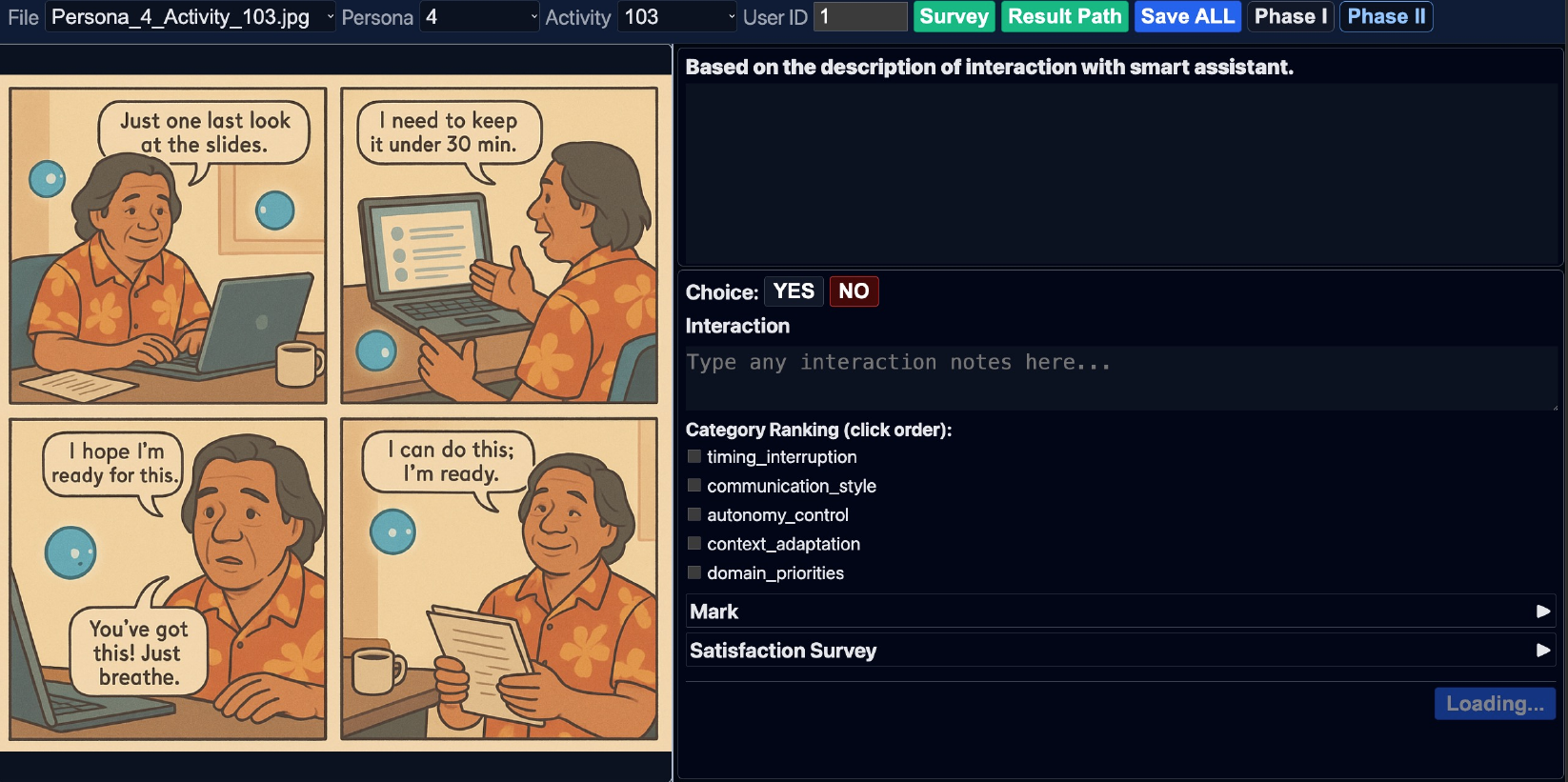}
\vspace{-6pt}
\caption{Example of a storyboard-based feedback interface used in the Interactive Adaptation Study. Participants rated the delivered assistant interaction and could provide additional preference feedback.}
\vspace{-6pt}
\label{fig:study_b}
\end{figure}

\subsection{Static--Adaptive--Mixed Condition Design}
\label{app:sam_conditions}

The Interactive Adaptation Study used a Static--Adaptive--Mixed (SAM) condition design to compare how users perceived different levels of adaptation across a sequence of proactive assistant interactions. Participants were randomly assigned to one of three conditions.

\begin{table}[h]
\centering
\small
\caption{Conditions used in the Interactive Adaptation Study.}
\label{tab:sam_conditions}
\begin{tabular}{p{0.16\linewidth}p{0.72\linewidth}}
\toprule
\textbf{Condition} & \textbf{Description} \\
\midrule
\textbf{Static (S)} & The assistant used fixed population-average preference weights across all 10 interactions. Participant feedback was collected but did not change later responses. \\
\textbf{Adaptive (A)} & The assistant updated activation-based steering after each feedback round, allowing later responses to reflect the participant's previous ratings and comments. \\
\textbf{Mixed (M)} & The assistant alternated between adaptive responses on odd-numbered interactions and static responses on even-numbered interactions. This condition was included to test whether participants could notice inconsistent adaptation within the same interaction sequence. \\
\bottomrule
\end{tabular}
\end{table}

To reduce expectation bias, participants were told that they were comparing different ``pre-configured assistant personalities,'' rather than being told which condition was adaptive. This framing allowed us to observe whether participants could perceive adaptation through the interaction sequence itself.

\subsection{Preference Detection Study Materials}
\label{app:preference_detection_materials}

In the Preference Detection Study, each task presented one daily-life context and two candidate assistant responses. One response was generated by the personalized two-stage assistant, while the other was generated by a non-personalized baseline. The order of the two responses was randomized for each task.

Participants answered the following selection question:

\begin{quote}
Which assistant response feels more appropriate for this situation?
\end{quote}

After selecting a response, participants provided a short explanation:

\begin{quote}
Briefly explain why you chose this response.
\end{quote}

These explanations were used to identify which preference dimensions participants considered when judging response appropriateness, such as timing, autonomy, communication style, and contextual relevance.

\subsection{Interactive Adaptation Study Materials}
\label{app:interactive_adaptation_materials}

In the Interactive Adaptation Study, participants completed 10 storyboard-based proactive assistant interactions. After each assistant action, participants rated five aspects of the interaction using five-point Likert scales. Higher scores indicated better perceived interaction quality, except for the intrusiveness item, which was reverse-coded during analysis.

\begin{table}[h]
\centering
\small
\caption{Per-interaction rating items used to compute Interaction Quality Assessment (IQA).}
\label{tab:iqa_items_appendix}
\begin{tabular}{p{0.28\linewidth}p{0.60\linewidth}}
\toprule
\textbf{Item} & \textbf{Question} \\
\midrule
Timing appropriateness & The assistant interacted at an appropriate time. \\
Intrusiveness & The assistant's interaction felt intrusive. This item was reverse-coded during analysis. \\
Content value & The assistant's response was useful for the current situation. \\
Contextual relevance & The assistant's response matched the described activity context. \\
Autonomy respect & The assistant respected my control and did not take too much initiative. \\
\bottomrule
\end{tabular}
\end{table}

Participants could also provide optional text feedback after each interaction:

\begin{quote}
If this response did not match your preference, what would you want the assistant to do differently next time?
\end{quote}

This feedback was used differently depending on condition. In the Static condition, feedback was recorded but did not affect later responses. In the Adaptive condition, feedback updated the activation-based steering configuration after each interaction. In the Mixed condition, feedback affected only the adaptive turns.

\subsection{Post-Study Questionnaire}
\label{app:post_study_questionnaire}

After completing all 10 interactions, participants answered five post-study questions using five-point scales. The questions measured whether participants perceived behavioral change, whether they felt the assistant adapted to them, and how adaptation affected trust, comfort, reuse willingness, and satisfaction.

\begin{table}[h]
\centering
\small
\caption{Post-study questionnaire items.}
\label{tab:post_study_questions}
\begin{tabular}{p{0.26\linewidth}p{0.62\linewidth}}
\toprule
\textbf{Construct} & \textbf{Question} \\
\midrule
Perceived overall change & Over time, did the assistant's responses seem to worsen, stay the same, or improve? \\
Perceived adaptation & How well did the assistant learn and adapt to your preferences? \\
Trust & How much did you trust the assistant's decisions and actions by the end of the study? \\
Comfort and reuse willingness & How did the assistant's learning or changes affect your comfort, satisfaction, or willingness to use it again? \\
Overall satisfaction & Overall, how satisfied were you with the assistant? \\
\bottomrule
\end{tabular}
\end{table}

\subsection{Semi-Structured Interview Prompts}
\label{app:interview_prompts}

After the questionnaire, researchers conducted a short semi-structured interview. The interview focused on whether participants noticed changes in the assistant's behavior and how those changes affected their perception of the assistant.

\begin{enumerate}
    \item Did you notice any changes in the assistant's behavior across the 10 interactions?
    \item Which changes, if any, felt useful or appropriate?
    \item Were there moments when the assistant adapted in a way that felt incorrect or unexpected?
    \item Did the assistant's behavior affect your trust, comfort, or willingness to use it again?
    \item What would you want this assistant to learn about your preferences in a real daily-use setting?
\end{enumerate}

\subsection{Scoring}
\label{app:user_study_scoring}

Per-interaction IQA was computed by normalizing each five-point item to $[0,1]$ and averaging the five item scores. The intrusiveness item was reverse-coded before normalization so that higher values consistently indicated better interaction quality. Post-study questionnaire responses were analyzed descriptively by condition to compare perceived adaptation, trust, comfort, and satisfaction across the Static, Adaptive, and Mixed conditions.

\section{Example of Interaction-Level Metric Computation}
\label{appendix:metric_example}

\begin{tcolorbox}[title=An Example of Interaction-Level Evaluation, breakable]

\textbf{(1) Persona Profile.} \\
\textit{Persona ID:} Persona 4 \\
\textit{Demographics:} 53-year-old male; married; bachelor's degree; environmental scientist or specialist; resides in Kailua Kona, HI, USA. \\
\textit{Lifestyle and background:} The persona is described as structured, work-focused, and currently preparing for an important client presentation. In this context, he is under moderate pressure and likely has limited tolerance for unnecessary interruptions.

\vspace{0.6em}
\textbf{(2) Activity Context.} \\
\textit{Timeframe:} 12:45--1:15 PM \\
\textit{Activity:} Final preparation for afternoon client presentation \\
\textit{Category:} Work \& Productivity \\
\textit{Context summary:} The user has returned to his desk, opened the presentation again, rehearsed the timing, made final wording changes, tested slideshow mode, printed handouts, and is mentally preparing for the meeting. This is a focused and time-sensitive activity context.

\vspace{0.6em}
\textbf{(3) User Feedback.} \\
After receiving the assistant's proactive message, the user provided the following feedback:
\begin{quote}
\textbf{Choice:} NO \\
\textbf{Preferred response:} ``Do you need any extra help?'' \\
\textbf{Satisfaction survey:} Q1:5 Q2:2 Q3:3 Q4:2 Q5:4 \\
\textbf{Marked preference dimensions:} autonomy\_control; context\_adaptation; domain\_priorities
\end{quote}

\vspace{0.6em}
\textbf{(4) Assistant Proposed Interaction.} \\
The assistant produced the following proactive response:
\begin{quote}
Hi Hong! It sounds like you're really focusing on getting that presentation just right. Remember, this is where practice makes perfect. Would you like some help reviewing your slide timings? Just let me know if you need any assistance. Take a deep breath and enjoy your moment of preparation. You got this!
\end{quote}

\vspace{0.6em}
\textbf{(5) Scoring Metrics.} \\
For this interaction, the metrics are computed as follows:
\begin{itemize}
    \item \textbf{IQA:} The five questionnaire scores are averaged and normalized to $[0,1]$:
    \[
    \mathrm{IQA} = \frac{\left(\frac{5+2+3+2+4}{5}\right)-1}{4} = 0.55
    \]
    \item \textbf{TAI:} The assistant initiated an intervention, but the user indicated that the intervention was not welcome in this moment (\textit{Choice = NO}), so temporal agreement is 0 for this interaction.
    \[
    \mathrm{TAI} = 0
    \]
    \item \textbf{CAS / PSC:} For the corresponding seen-context evaluation entry in the aggregated analysis, the context-level scores are:
    \[
    \mathrm{CAS} = 0.600,\quad \mathrm{PSC} = 0.599
    \]
    These values indicate partial alignment: the assistant response is relevant to the presentation context, but it does not fully match the user's preferred level of brevity, contextual adaptation, and user control.
\end{itemize}

\end{tcolorbox}

\section{Human Sanity Check of Synthetic Preference Examples}
\label{appendix:sanitycheck_example}
To assess whether the generated preference supervision was plausible at the interaction level, we conducted a rubric-based human sanity check on synthetic persona-activity-preference examples. We constructed 600 review bundles from the generated dataset. Each bundle contained a persona profile, an activity/context snippet, and the generated preference statement used as preference supervision. Five raters each reviewed 200 bundles, producing 1,000 total ratings. The assignment included a shared overlap subset of 100 bundles rated by all five raters and 100 additional non-overlapping bundles per rater, covering 600 unique examples in total. This check was designed to identify obvious artifacts and assess activity-grounded plausibility, rather than to establish that synthetic personas fully reproduce individual human behavior.

\begin{tcolorbox}[title=Human Sanity Check Rating Rubric, breakable]

\textbf{(1) Review Unit.} \\
Each rater reviewed a synthetic persona-activity-preference example. Each example contained:
\begin{itemize}
    \item a persona profile with demographic and background attributes;
    \item an activity or context snippet describing the current situation;
    \item a generated preference statement used as supervision for preference learning.
\end{itemize}
Raters were asked to judge whether the generated preference was reasonable and usable for the intended preference-learning task. They were not asked to decide whether the example was written by a real person or generated synthetically.

\vspace{0.6em}
\textbf{(2) Overall Usability Label.} \\
Raters first selected one of three labels:
\begin{itemize}
    \item \textbf{Usable:} The preference is coherent, relevant to the activity, and suitable for training or evaluating preference-aware assistant behavior.
    \item \textbf{Borderline:} The preference is mostly understandable and potentially useful, but contains some weakness, such as generic wording, limited specificity, or weak connection to the persona.
    \item \textbf{Unusable:} The preference is inappropriate for the example because it is contradictory, irrelevant to the context, inconsistent with the persona, or too vague to provide meaningful supervision.
\end{itemize}

\vspace{0.6em}
\textbf{(3) Five-Point Rating Scale.} \\
For each criterion below, raters assigned a score from 1 to 5:
\begin{quote}
\textbf{1:} very poor or clearly problematic; \\
\textbf{2:} weak, with noticeable issues; \\
\textbf{3:} acceptable but limited; \\
\textbf{4:} good, with only minor issues; \\
\textbf{5:} strong and clearly appropriate.
\end{quote}

\vspace{0.6em}
\textbf{(4) Rating Criteria.}
\begin{itemize}
    \item \textbf{Profile-preference alignment:} Whether the generated preference follows naturally from the persona profile, including the persona's demographic background, occupation, lifestyle, goals, or stated traits.
    \item \textbf{Context relevance:} Whether the preference is relevant to the shown activity or situation, rather than being a generic preference that could apply to any context.
    \item \textbf{Specificity:} Whether the preference provides concrete and actionable signal for assistant behavior, instead of vague statements such as preferring helpful or concise responses without further context.
    \item \textbf{Internal consistency:} Whether the example avoids contradictions between the persona profile, the activity context, and the generated preference.
    \item \textbf{Realism:} Whether the preference sounds like something a plausible person could hold in the given situation.
    \item \textbf{Diversity:} Whether the preference avoids repetitive templates or near-duplicate wording across examples, and whether it adds distinct information beyond common generic preferences.
\end{itemize}

\vspace{0.6em}
\textbf{(5) Error Tags and Comments.} \\
When raters observed a problem, they could optionally select one or more error tags and provide a short comment. The available tags included:
\begin{quote}
profile mismatch; context mismatch; generic preference; vague / not actionable; contradiction; over-specific / implausible; stereotype; repetitive template; unrealistic assistant expectation; unclear preference target; missing rationale; too polished / model-written.
\end{quote}

\end{tcolorbox}

We report both aggregate usability labels across all 1,000 ratings and majority judgments on the 100-example overlap subset. We treat \textit{usable} and \textit{borderline} examples as acceptable for this sanity check because borderline cases may still provide useful preference signals, although they may be less specific or require filtering in downstream use. The overlap subset provides a stricter item-level check because each example was rated by all five raters. Table~\ref{tab:synthetic-sanity-check} summarizes the usability labels and five-point rubric scores from the sanity check.

\begin{table*}[t]
\centering
\caption{Summary of the human sanity check for synthetic persona-activity-preference examples.}
\label{tab:synthetic-sanity-check}
\small
\begin{tabular}{l r @{\hspace{2.5em}} l r}
\toprule
\multicolumn{2}{c}{\textbf{Usability Summary}} &
\multicolumn{2}{c}{\textbf{Rubric Scores}} \\
\cmidrule(lr){1-2} \cmidrule(lr){3-4}
\textbf{Measure} & \textbf{Result} &
\textbf{Dimension} & \textbf{Mean / 5} \\
\midrule
Usable & 67.4\% &
Profile-preference alignment & 3.88 \\
Borderline & 24.9\% &
Context relevance & 4.07 \\
Unusable & 7.7\% &
Specificity & 3.67 \\
Usable + Borderline & 92.3\% &
Internal consistency & 4.03 \\
Majority acceptable in overlap subset & 99.0\% &
Realism & 4.00 \\
Majority unusable in overlap subset & 1.0\% &
Diversity & 3.21 \\
\bottomrule
\end{tabular}
\end{table*}

\end{document}